\documentclass[a4paper,fleqn,usenatbib]{mnras}

\usepackage{newtxtext,newtxmath}

\usepackage[T1]{fontenc}
\usepackage{ae,aecompl}

\usepackage{graphicx}	
\usepackage{amsmath}	
\usepackage{amssymb}	
\usepackage{soul}

\usepackage{longtable}
\usepackage{threeparttablex}
\usepackage{booktabs}
\usepackage{epstopdf}
\usepackage{times}       
\usepackage{float}

\title[Planetary signatures around white dwarfs]{The unbiased frequency of planetary signatures around single and binary white dwarfs using {\it Spitzer} and {\it Hubble}}

\author[T. G. Wilson et al.]{
Thomas G. Wilson$^{1,2}$\thanks{E-mail: tgw@star.ucl.ac.uk},
Jay Farihi$^{1}$,
Boris T. G{\"a}nsicke$^{3}$,
and Andrew Swan$^{1}$
\\
$^{1}$Department of Physics \& Astronomy, University College London, London WC1E 6BT, UK\\
$^{2}$Isaac Newton Group, Apartado 321, E-38700 Santa Cruz de La Palma, Spain \\
$^{3}$Department of Physics, University of Warwick, Coventry CV4 7AL, UK
}

\pubyear{2019}

\begin{document}
\label{firstpage}
\pagerange{\pageref{firstpage}--\pageref{lastpage}}
\maketitle

\begin{abstract}

This paper presents combined {\em Spitzer} IRAC and {\em Hubble} COS results for a double-blind survey of 195 single and 22 wide binary white dwarfs for infrared excesses and 
atmospheric metals. The selection criteria include cooling ages in the range 9 to 300\,Myr, and hydrogen-rich atmospheres so that the presence of atmospheric metals
can be confidently linked to ongoing accretion from a circumstellar disc. The entire sample has infrared photometry, whereas 168 targets have corresponding ultraviolet spectra. Three stars 
with infrared excesses due to debris discs are recovered, yielding a nominal frequency of $1.5_{-0.5}^{+1.5}$\,per cent, while in stark contrast, the fraction
of stars with atmospheric metals is $45\pm4$\,per cent. Thus, only one out of 30 polluted white dwarfs exhibits an infrared excess at 3--4\,$\upmu$m in IRAC photometry, which reinforces the 
fact that atmospheric metal pollution is the most sensitive tracer of white dwarf planetary systems. The corresponding fraction of infrared excesses around white
dwarfs with wide binary companions is consistent with zero, using both the infrared survey data and an independent assessment of potential binarity for well-established dusty and 
polluted stars. In contrast, the frequency of atmospheric pollution among the targets in wide binaries is indistinct from apparently single stars, and moreover
the multiplicity of polluted white dwarfs in a complete and volume-limited sample is the same as for field stars. Therefore, it appears that the delivery of planetesimal material onto 
white dwarfs is ultimately not driven by stellar companions, but by the dynamics of planetary bodies.

\end{abstract}




\section{Introduction}
\label{sec:SEC1}

Despite the myriad exoplanetary systems known to exist via transit and radial velocity surveys, knowledge of their 
chemistry and assembly remains limited. Studies of protoplanetary discs reveal chemical signatures of the solids and gases 
that precede and are likely incorporated into planetesimals and eventually planets \citep{Bergin,Marty}, while exoplanet 
detection provides the sizes (and sometimes masses) and architectures for a limited set of end results. Direct imaging 
has demonstrated that wide and massive giant planets are not as frequent as those detected by transit and radial velocity methods, yet 
these techniques are sensitive only to a fraction of Solar system like architectures \citep{Gillon}. The detection of an Earth-like exoplanet around a Sun-like star may be on the horizon, but information on the composition is critical and cannot be provided by conventional means.  

From an observational perspective, it is now well established that planetary systems can survive into the post-main sequence.
Many white dwarf stars exhibit atmospheric metals that should otherwise sink (\citealt{Zuckerman2003}; \citealt*{Koester2014}),
tens of these display infrared excesses consistent with planetary materials (\citealt{Reach}; \citealt*{Jura2009}), and at
least one system has complex transiting events \citep{Vanderburg}. These planetary systems orbiting white
dwarfs universally exhibit atmospheric metal pollution, where closely-orbiting circumstellar discs -- the source of the 
atmospheric heavy elements -- are often detected via dusty and gaseous emission (see \citealt{Farihi2016} and references therein).
Importantly, white dwarf atmospheres distil the disc material and provide an indirect, but observable elemental composition 
of the parent body or bodies (see \citet{Jura2014} and references therein). To date, the observed abundances in polluted
white dwarfs have been dominated by objects with strikingly Earth-like chemistry \citep{Wilson}, but with substantial
diversity (\citealt{Jura2015}; \citealt{Kawka}; \citealt*{Hollands2018a}), including objects relatively rich in water, but poor in 
other volatiles (\citealt*{Farihi2013a}; \citealt{Raddi}), and a single instance of an ice-rich body \citep{Xu2017}. The nature
and frequency of these planetary systems therefore plays a critical and complementary role to the study of the 
exoplanetary systems via other methods such as transit photometry, precision radial velocities, and direct imaging. No 
planetary system orbiting a main-sequence star can provide this type of compositional information, and thus polluted
white dwarfs are key to understanding the formation and evolution of planetesimals and their associated planets. 

Theoretical studies have demonstrated that planet-planetesimal perturbations can create highly eccentric orbits that deliver 
substantial minor planetary body masses to the innermost orbital regions around white dwarf stars (\citealt{Veras2013}; \citealt*{Mustill}). 
In some cases, the dynamical interactions are favourable to generating planetesimal orbits with periastra interior to the stellar 
Roche limit (\citealt*{Debes2012}; \citealt{Frewen}), thus leading to tidal fragmentation.  In other models, namely those with 
bodies originating in a Kuiper-like belt, minor body perturbations by a single planet are insufficient to create such star-grazing 
orbits, and additional (planetary) gravitational encounters are necessary \citep{Bonsor2012}. 

In many cases, the delivery of planetary material into the innermost region and eventually onto the white dwarf surface requires full-fledged planets, but their detection has so 
far been elusive due to insufficient sensitivity \citep{Mullally}. The formation of the observed circumstellar discs is an area of
ongoing study, where collisions are likely to play a role owing to the fact that the shrinkage of wide and highly eccentric orbits 
by Poynting-Robertson (PR) drag requires $10^5-10^6$\,yr or longer for millimetre and centimetre-size particles, respectively 
\citep{Veras2015}. Once formed, a collision-less disc of solids should evolve via PR drag \citep{Rafikov2011a,Bochkarev},
whether optically thick or optically thin to starlight. If collisions are important or there is significant gas present that is
co-spatial with the dust, then competing effects may drive the disc evolution (\citealt{Rafikov2011b}; \citealt*{Metzger,Kenyon}).

A prediction made by dynamical models is that the frequency of white dwarf pollution and circumstellar debris discs should be 
modestly or strongly time-dependent, based on the nature of the instability that drives material inward (see and compare 
e.g. \citealt{Debes2012,Mustill}; \citealt*{Petrovich}; \citealt{Smallwood}). Thus, the empirical occurrence rate for atmospheric 
metals and infrared excesses are necessary to test these models. While the frequency of white dwarfs with photospheric metals 
has been robustly determined to be at least 20--30\,per cent \citep{Zuckerman2010,Koester2014}, the bulk of published {\em Spitzer} 
searches for infrared excesses have targeted stars with known atmospheric metals in order to formalise the link between discs and 
pollution (\citealt*{Jura2007}; \citealt{Farihi2010}). To date, the frequency of debris discs that are detected in the infrared has been 
constrained to be between 1 and 4\,per cent (\citealt{Mullally}; \citealt*{Farihi2009b}; \citealt{Barber,Rocchetto}).  
However, the bulk of published work has suffered from selection biases, sensitivity issues, or insufficient statistics (or 
a combination of these). Wide-field surveys such as the Sloan Digital Sky Survey and {\em WISE} have been utilised to characterise infrared 
excess frequency using the largest possible number of stars, but can suffer from limited sensitivity and poor sample characterisation 
\citep{Debes2011,Girven}. Therefore, an unbiased {\em Spitzer} survey is needed to accurately measure the occurrence of discs that exhibit infrared excesses.

This study analyses {\em Spitzer} IRAC observations of hydrogen-rich atmosphere white dwarfs that were selected on the 
basis of cooling age, and thus without any bias toward the potential presence of atmospheric metals. The same stars were part 
of several {\em Hubble} COS Snapshots, thereby providing a double-blind study of metal pollution and infrared excess frequency. 
These observations form the largest available, unbiased {\em Spitzer} sample to assess infrared excess and atmospheric metal 
frequencies, and form the only such double-blind study. The {\em Spitzer} observations are presented in Section~\ref{sec:SEC2}, 
with the data analysis reported in Section~\ref{sec:SEC3}. The results and notes on individual targets are discussed in Section~\ref{sec:SEC4}, 
with conclusions given in Section~\ref{sec:SEC5}. The results of the {\em Hubble} Snapshots will be discussed in a forthcoming paper.

\section{Observations and Data}
\label{sec:SEC2}

\begin{figure}
\includegraphics[width=\linewidth]{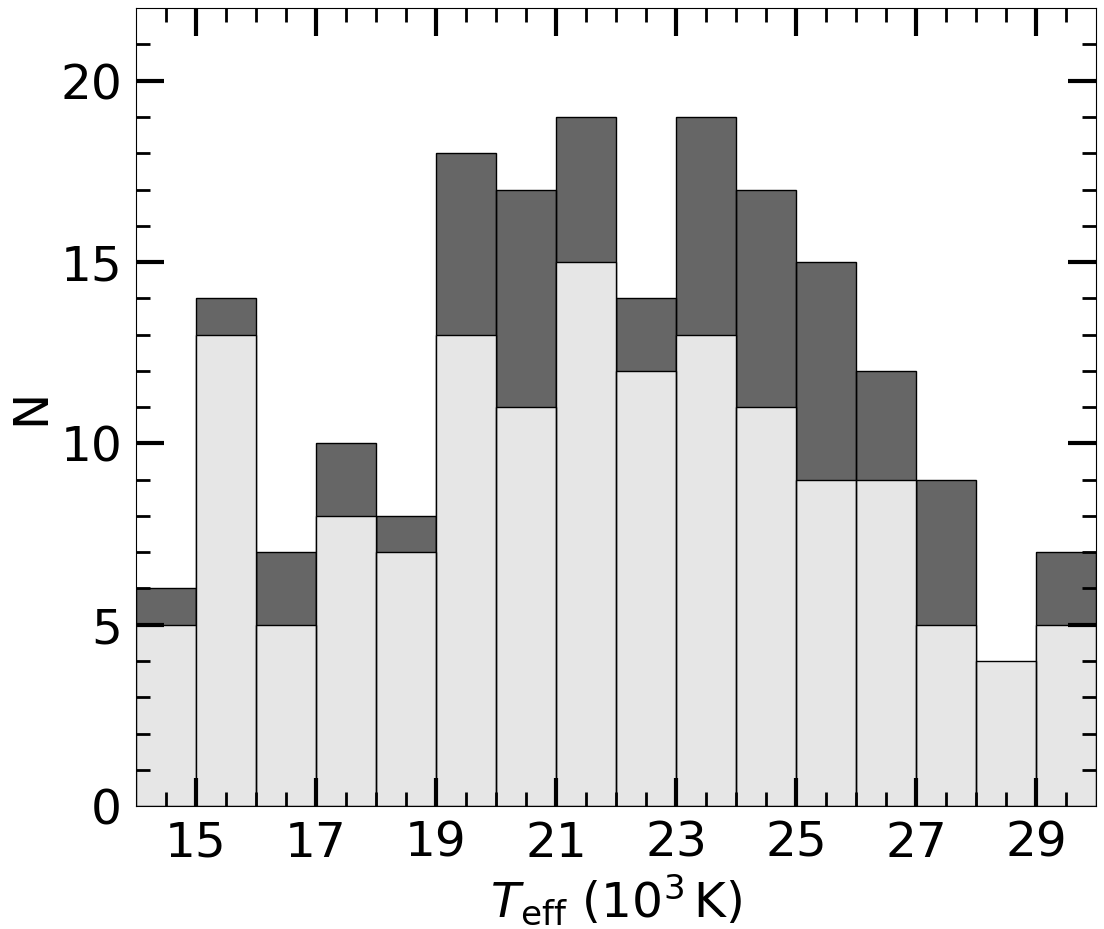}
\caption{$T_{\rm eff}$ histogram for the single white dwarfs in this study. All stars were observed with {\it Spitzer} and are shown in 
dark grey, while the subset observed by {\it Hubble} are shown in light grey. The same number of targets were observed with both telescopes in one of the bins.}
\label{fig:FIG1}
\end{figure}

\subsection{Sample of single and binary white dwarfs}
\label{sec:SEC2.1}

This paper reports observations that are one part of a dual instrument survey to identify both atmospheric metal and infrared 
excess frequencies for white dwarfs as a function of effective temperature and hence cooling age. All targets and their associated stellar parameters in the study were taken 
from catalogues of nearby white dwarfs (\citeauthor*{Liebert} \citeyear{Liebert}; \citeauthor{Koester2009a} \citeyear{Koester2009a};
\citeauthor*{Gianninas} \citeyear{Gianninas}). Sample stars have been selected to have hydrogen-rich atmospheres, owing to their relatively 
short diffusion timescales \citep{Koester2009b}, in order to confidently ascribe any atmospheric metals to ongoing accretion from 
circumstellar material. Another criterion for each target is $F_{\uplambda}(1300\,$\AA$)>5\times10^{-14}$\,erg\,cm$^{-2}$\,s$
^{-1}$\,\AA$^{-1}$ to achieve a sufficiently high S/N in the far ultraviolet COS spectra. These requirements result in a 
range of stellar effective temperatures 14\,000--31\,000\,K, and thus probe post-main sequence planetary system dynamics at ages of 9--300\,Myr.

The selection criteria above result in 236 sources. An extensive literature and database search indicates that 40 targets are confirmed or candidate binary systems of varying composition and separation. The remaining 196 apparently single white dwarfs form the basis for the bulk of the study. The binary subsample is discussed separately, as the sample statistics are less robust. Table~\ref{tab:TABA1} lists all stars for which IRAC photometry was attempted, with binaries noted. 

\subsection{{\em Spitzer} observations}
\label{sec:SEC2.2}

Imaging data for the entire sample were obtained with the Infrared Array Camera (IRAC; \citeauthor{Fazio} \citeyear{Fazio}) onboard the {\em Spitzer Space 
Telescope} \citep{Werner} and are the main focus of this study. The same sample was part of multiple {\em Hubble} Snapshot surveys 
using COS, of which 168 targets were observed (cf. 236 observed by {\it Spitzer}), with the details to be reported in a subsequent publication. 
New observations for 168 white dwarfs were taken with warm {\em Spitzer} IRAC in Cycles 8 and 12 for Programs 80149 and 12103, respectively. Targets were observed at both 
3.6 and 4.5\,$\upmu$m, where images were taken with 20 frames of 30\,s each using the medium-sized, cycling dither pattern, and 
resulting in a total exposure time of 600\,s in each warm IRAC channel. Archival (cryogenic or warm) IRAC observations for an additional 68 sources were retrieved and analysed to complete the sample.
As an indication of the survey overlap between instruments, the number of single white dwarfs observed by {\em Spitzer} and {\em Hubble} as a function of effective temperature is presented in Fig.~\ref{fig:FIG1}.  

\subsection{IRAC photometry}
\label{sec:SEC2.3}

Single, fully processed, and calibrated 0.6\,arcsec pixel$^{-1}$ mosaic images were extracted for all targets in the observed 
bandpasses by the IRAC calibration pipeline S19.2.0. Aperture photometry was conducted using the standard {\sc iraf} task 
{\sc apphot} with aperture radii of 2.4 or 3.6\,arcsec, depending on target brightness and additional nearby sources, and 
14.4--24.0\,arcsec sky annuli. The fluxes were corrected for aperture size using conversion factors listed in the IRAC 
Data Handbook, but not corrected for colour. For all targets with sufficiently bright neighbouring sources (including binaries), where photometric contamination was possible or likely, PSF-fitting photometry was conducted using {\sc apex}. 
The error in the measured flux was summed in quadrature with the calibration uncertainty that was taken to be 5\,per cent \citep*{Farihi2008}.  
The newly measured fluxes and corresponding uncertainties 
for the entire sample are presented in Table~\ref{tab:TABA1}. To calculate upper limits, 
aperture photometry was conducted at the published coordinates using a 2.4\,arcsec radius. After aperture correction the 
determined flux within the aperture was compared to the sky noise per pixel multiplied by the aperture area, with the larger 
value reported as the upper limit. One target (1339+346) was irreversibly contaminated by a nearby background source, and is not considered further.

It has been shown that intra-pixel variations in the {\em Spitzer} IRAC detectors can alter the measured flux by a few per cent, 
especially at 3.6\,$\upmu$m, which could affect the calculated excesses \citep*{Mighell}. The pixel-phase response can be 
modelled by a 2D Gaussian that is offset from the centre of the pixel, with the observed flux variation being determined by the position 
of the point-spread function (PSF) peak in the pixel. In order to test that the observed fluxes are robust against intra-pixel flux 
variation, simulations were conducted. Fluxes were calculated by modelling the pixel-phase response at 20 random intra-pixel positions.
The average flux variation was determined to be 0.1\,per cent. Therefore, with well-dithered observations the flux variation in the 
analysed mosaicked frames is negligible and errors on the report fluxes will be dominated by measurement and calibration uncertainties.

Spectral energy distributions (SEDs) were constructed for all stars using additional photometric data from various catalogues including: 
AAVSO Photometric All-Sky Survey (\citealt{Henden}), Deep Near Infrared Survey of the 
Southern Sky (\citealt{Epchtein}), Panoramic Survey Telescope and Rapid Response System 
(\citealt{Chambers}), Sloan Digital Sky Survey (\citealt{Ahn}), Two Micron All Sky 
Survey (\citealt{Cutri}), and UKIRT Infrared Deep Sky Survey (\citealt{Lawrence}). Near-infrared photometric data for several stars were taken from the literature \citep{Farihi2009a,Barber}. 
White dwarf atmospheric models \citep{Koester2010} were fitted to the available optical through near-infrared fluxes using a least-squares algorithm. 

For each star in the sample, model fitting to the available data was conducted starting near the literature value of $T_{\rm eff}$, and varying up to $\pm2000$\,K in steps of 500\,K. 
The temperature that resulted in the smallest fitting error was adopted and was typically within 1000\,K of the literature value. 
On average, for a 500\,K deviation from the best fit temperature, the model fitting error increased by 7\, per cent of the minimal value.

\section{Data Analysis}
\label{sec:SEC3}

The following analysis applies only to those single and binary targets where it is possible to recover reliable photometry of the white dwarf. There are 210 such science targets; 195 of these are apparently single, and 15 in spatially-resolved binaries. Those binary systems whose IRAC photometry is dominated or influenced by unresolved or marginally resolved companions are not searched for infrared excesses. However, for many such systems, corresponding COS observations in the far-ultraviolet will effectively probe isolated white dwarfs. 

In order to comprehensively establish the frequency of detectable debris discs, two infrared excess determination methods 
are used: the flux method and the colour method. For each of these methods discussed below, the photospheric model flux 
is determined by performing synthetic photometry on the atmospheric model over the IRAC bandpasses, with an excess considered significant if it is greater than 3$\upsigma$. 
To assess how sensitive the observations are in discovering a debris disc, the minimum detectable excess, $\eta$, is calculated for each star. 
Assuming that excesses are produced by a flat, opaque disc with an inner radius corresponding to $T_{\rm in} = 1400$\,K and $r_{\rm out} = {\rm R_{\odot}}$,
the $\eta$ value can be used to determine the inclination, $i$, at which a debris disc is no longer observable \citep{Bonsor2017}. 
By calculating the average $\eta$ values in each warm IRAC channel it is determined that all discs with $i < 89^{\circ}$ should be detected.

\subsection{Method 1: Flux excess}
\label{sec:SEC3.1}

Previous {\it Spitzer} work searching for warm dust around 180\,000 {\it Kepler} field stars has used the flux excess method 
\citep{Kennedy}. In this method, the observations are compared to the model predictions for all IRAC channel measurements using the following statistic.

\begin{equation}
\upchi = \frac{F_{\rm obs} - F_{\rm mod}}{\sqrt{\upsigma^2_{\rm obs} + \upsigma^2_{\rm mod}}}
\label{eq:EQ1}
\end{equation}

\medskip
\noindent
Here $F_{\rm obs}$ and $F_{\rm mod}$ are the observed and model fluxes, respectively, $\upsigma_{\rm obs}$ represents the 
photometric measurement and calibration uncertainties summed in quadrature and are typically 5\,per cent of $F_{\rm obs}$. 
Errors in the model fitting to the short wavelength data are given as $\upsigma_{\rm mod}$ and are typically 1\,per cent of $F_{\rm mod}$.

\begin{figure}
\includegraphics[width=\linewidth]{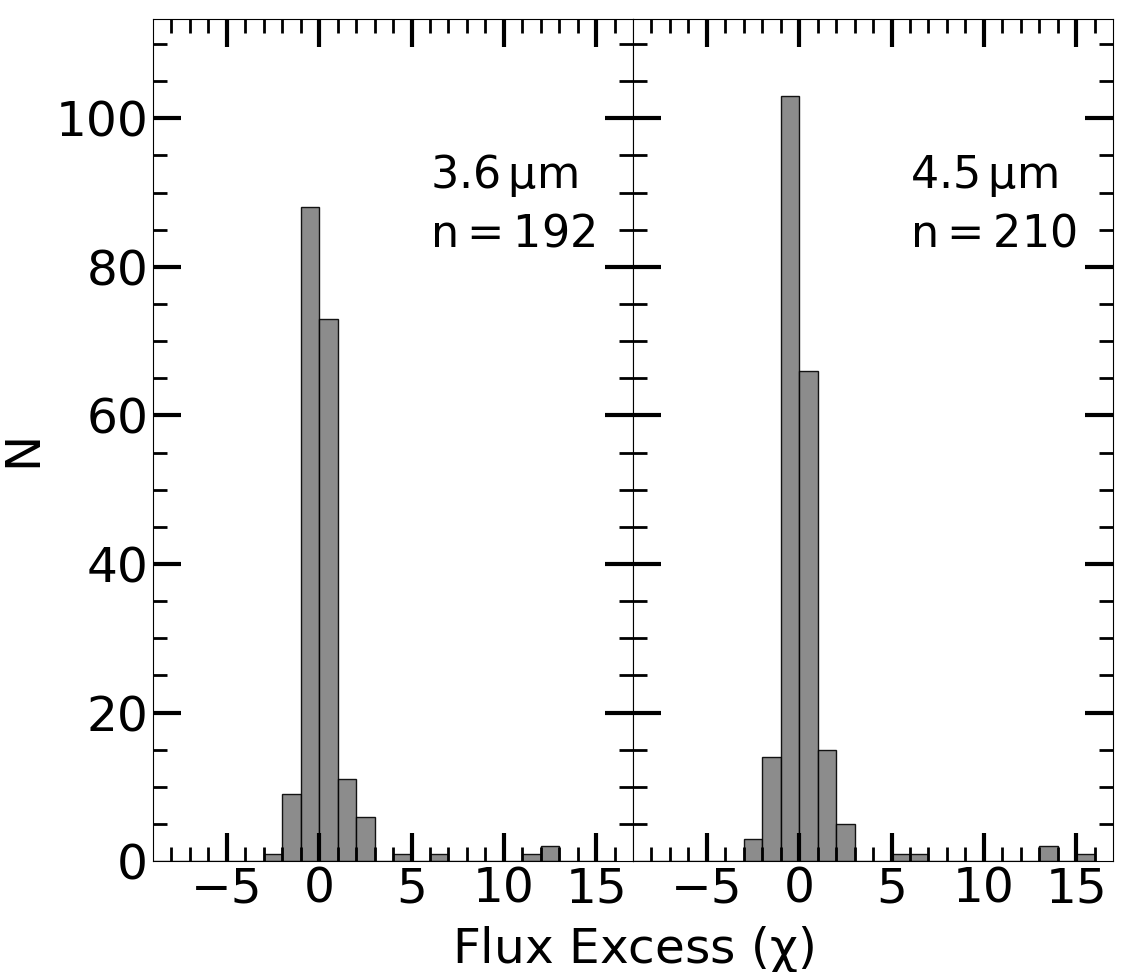}
\caption{Histograms of flux excess values, $\upchi$, as determined by Equation~(\ref{eq:EQ1}) for both warm IRAC channels. 
It should be noted that not all stars have 3.6\,$\upmu$m data, as some archival targets only have 4.5 and 7.9\,$\upmu$m observations.}
\label{fig:FIG2}
\end{figure}

Flux excesses, denoted by $\upchi$, are calculated using Equation~(\ref{eq:EQ1}), and are plotted
as histograms for each warm IRAC channel in Fig.~\ref{fig:FIG2}. For this sample, the mean and standard deviation in the flux excess
statistic are $\langle \upchi \rangle_{\rm 3.6} = 0.15\pm1.07$ and $\langle \upchi \rangle_{\rm 4.5} = 0.05\pm1.15$. These are broadly consistent with Gaussian distributions. 

Infrared excess candidates are identified by $\upchi > 3\upsigma$ in either IRAC bandpass. There are five stars with previously discovered infrared excesses: 0843+516, 1015+161, 1018+410, 1457--086, and 2328+107. 
All are recovered as significant using this analysis, with $\upchi$ values in the range 5--15$\upsigma$, as can be seen in Table~\ref{tab:TAB1}. Of the remaining sources, 
two systems exhibit $\upchi > 3\upsigma$ in both warm IRAC channels and are also reported in Table~\ref{tab:TAB1}. These two sources -- 1132+470 and 2218--271 -- are discussed in Section~\ref{sec:SEC4.3}.

\subsection{Method 2: Colour excess}
\label{sec:SEC3.2}

Infrared excess determination via colour can potentially reveal faint debris discs that might be missed using the flux
excess method. Examples of excesses calculated on the basis of a single colour, or weighted multiple colours, have been used 
to detect faint infrared excesses towards main-sequence stars using {\it WISE} photometry (\citealt*{Patel2014}; \citealt{Patel2017}).

In the following method, the observed and model fluxes, and their corresponding errors have been converted into (Vega) 
magnitudes. The IRAC colour excesses are determined using the following, with all quantities in units of magnitude.

\begin{equation}
\upSigma_{\rm ij} = \frac{m_{\rm i,obs} - m_{\rm j,obs} - m_{\rm ij,mod}}{\sqrt{\upsigma^2_{\rm i,obs} + \upsigma^2_{\rm j,obs} + \upsigma^2_{\rm ij,mod}}}
\label{eq:EQ2}
\end{equation}

\medskip
\noindent
Here $i$ and $j$ represent the two bandpasses that determine the colour, therefore the numerator in the above equation is the difference 
between the measured and model colour. The uncertainties in the observed magnitudes are added in quadrature with the uncertainty 
in the model colour. If sources have infrared photometry in three or more bandpasses, it is possible to construct an excess statistic
based on a weighting of the various colours. However, because fewer than 20\,per cent of the sample have observations at either 5.7 or 7.9\,$\upmu$m, 
and given that the longer wavelength arrays are substantially less sensitive \citep{Fazio}, these data are not used in the overall assessment. 
Therefore, this study is restricted primarily to data in the 3.6 and 4.5\,$\upmu$m bandpasses, and thus only a single colour excess statistic is used.

\begin{figure}
\includegraphics[width=\linewidth]{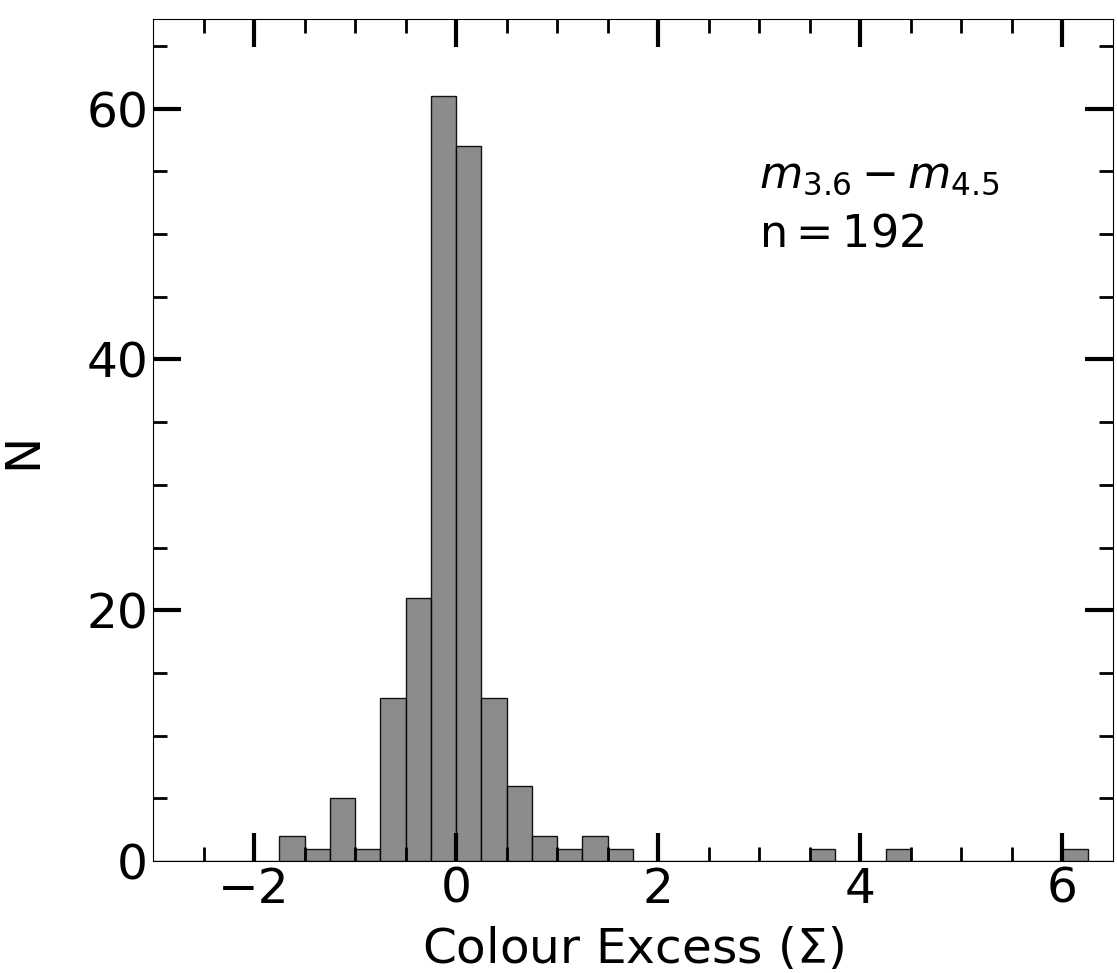}
\caption{Histogram of $m_{\rm 3.6}-m_{\rm 4.5}$ colour excess values, $\upSigma$, as determined by Equation~(\ref{eq:EQ2}) for stars observed in both warm IRAC channels.}
\label{fig:FIG3}
\end{figure}

Colour excesses, denoted by $\upSigma$, are calculated for all white dwarfs observed with {\em Spitzer} in both warm IRAC channels
using Equation~(\ref{eq:EQ2}), and a histogram of these $m_{\rm 3.6}-m_{\rm 4.5}$ colour excesses is shown in Fig.~\ref{fig:FIG3}. 
The mean and standard deviation are $\langle \Sigma \rangle = 0.02\pm0.79$.

Candidates are identified by $\upSigma\geq3\upsigma$, where three stars meet this criterion: 0843+516, 1015+161, 
and 1018+410. It is immediately apparent that this technique missed four stars that are flagged by the flux excess method (1132+470,
1457--086, 2218--271, and 2328+107; see Section~\ref{sec:SEC4.3}). The $\upSigma$ values for these stars are reported in Table~\ref{tab:TAB1}. 
This method may be powerful when multiple infrared bandpasses are available, as this adds information that is not present with only IRAC 3.6 and 4.5\,$\upmu$m 
photometry. As it stands, for the subsample with data in both channels, the flux excess method is more sensitive, especially to SEDs 
that deviate modestly from a Rayleigh-Jeans slope. Interestingly, only white dwarfs with $\upchi>10\upsigma$ are seen to have a corresponding $\upSigma>3\upsigma$ as shown in Fig.~\ref{fig:FIG4}, 
but this can be partly accounted for by the additional uncertainty of utilising two measurements for colour excesses \citep{Patel2014}. 
Although the stars flagged by the colour excess method are fewer, the results in Section~\ref{sec:SEC4} demonstrate they are a higher fidelity sample of bona fide discs.

\subsection{Atmospheric metals}

{\it Hubble} COS data were obtained for a subsample of stars that also have {\it Spitzer} IRAC measurements. The 
analysis of these ultraviolet spectra requires detailed atmospheric modelling, including the effects of radiative levitation, and is 
beyond the scope of this paper. A subset of the {\em Hubble} spectra has been published in \cite{Koester2014}, and the full 
dataset is the subject of ongoing work. This study focuses on the frequency of discs detectable in the infrared with {\em Spitzer}, 
but a fundamental issue is sensitivity to circumstellar material, as it is established that many metal-enriched stars lack 
infrared excesses \citep{Rocchetto, Farihi2016}. For this reason, a visual inspection of the {\em Hubble} spectra is conducted, 
with the absence or presence of Si\,{\sc ii} noted. Among 168 observed stars, 78 exhibit photospheric silicon.

\begin{table}
\centering
\caption{Infrared excess statistics for stars with values above $3\upsigma$ using at least one method. Columns are named after 
the wavelength in microns at which the excesses are determined.}
\begin{tabular}{crrc}

\hline    
     
&\multicolumn{2}{c}{......... $\upchi$ .........} 	&......... $\upSigma$ ......... \\

WD 	            &(3.6) 	&(4.5) 	&($m_{\rm 3.6}-m_{\rm 4.5}$) \\ 

\hline
     
0843+516 		&12.9 	&15.4 	&6.1\\ 		
1015+161 		&11.8 	&13.7 	&3.8\\ 		
1018+410 		&12.3 	&13.2 	&4.4\\ 		
1132+470 		&7.6 	&7.3 	&0.0\\ 		
1457--086 	    &6.0 	&6.0 	&0.2\\ 	
2218--271 	    &6.2 	&7.9 	&1.9\\ 	
2328+107 		&4.9 	&5.5 	&0.6\\ 	

\hline
      
\label{tab:TAB1} 
\end{tabular}
\end{table}

\section{Results}
\label{sec:SEC4}

Below, the results are presented separately for those apparently single white dwarfs versus those confirmed or suspected to be in binary systems.

\subsection{Single white dwarfs}
\label{sec:SEC4.1}

Using the methods outlined above, out of the 195 nominally single white dwarfs with good IRAC photometry, seven systems display a significant infrared excess.
However, as discussed in more detail below, not all of the excesses are thought to be due to planetary debris discs. 
There are three potential causes of infrared excesses: circumstellar debris discs, binary companions, or background contamination. 
This subsection focuses on three excesses thought to be due to debris discs, while Section~\ref{sec:SEC4.3} discusses three systems whose excesses are possibly consistent with binarity and one with contamination.

\begin{figure}
\includegraphics[width=\linewidth]{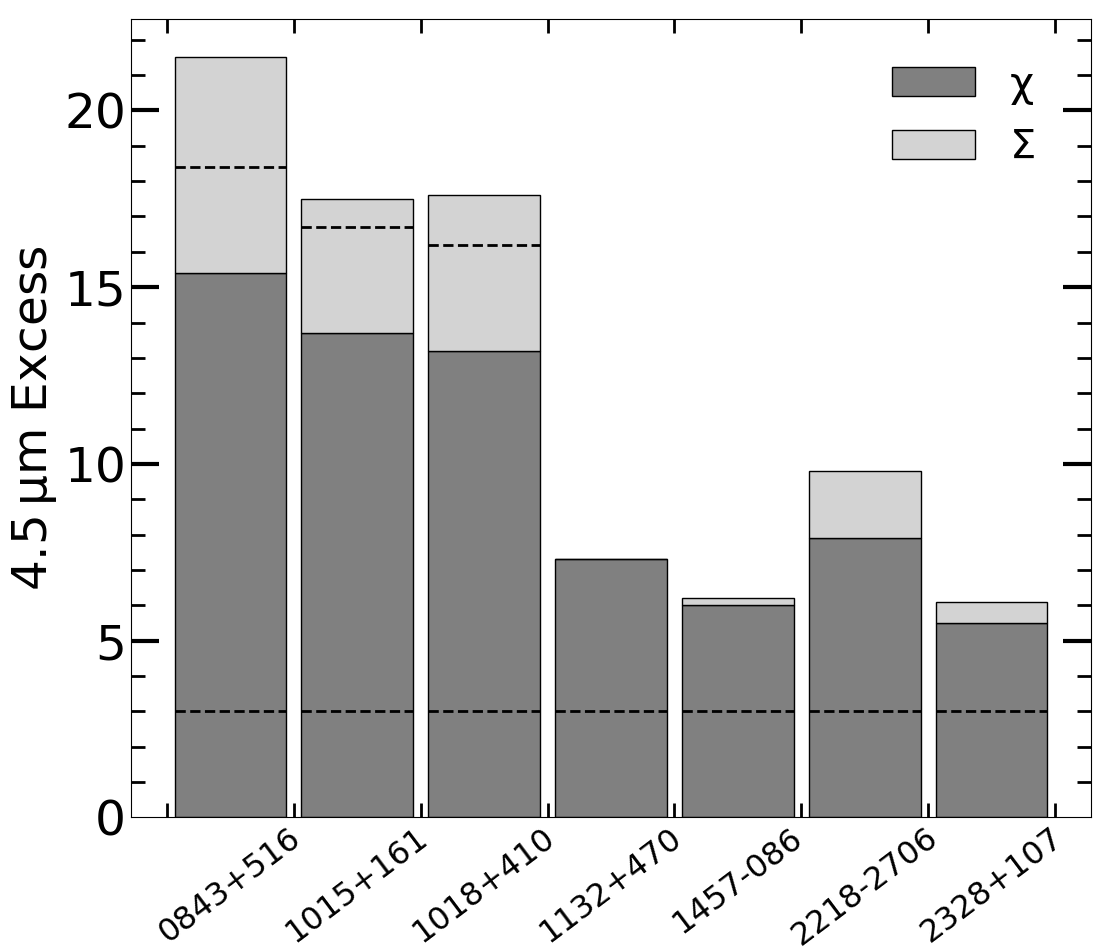}
\caption{Photometric excesses at 4.5\,$\upmu$m as determined by both methods. The black dashed lines 
represent $3\upsigma$ for each method. Only three stars with bona fide debris discs have a significant infrared excess using both methods.}
\label{fig:FIG4}
\end{figure}

These stars and excess values are reported in Table~\ref{tab:TAB1}.
Interestingly, it should be noted that the three systems with an excess due to a planetary debris disc have significant excesses using both methods, 
whereas the objects with an excess that can be explained by other causes only have a significant excess using the flux method. 
This apparent link between bona fide discs and significant excesses in both methods can be seen clearly in Fig.~\ref{fig:FIG4}. 
Furthermore, by calculating the excesses of known dusty white dwarfs using both methods it can be seen that 92\,per cent have a significant excess using both the flux and colour methods.

The following analysis is restricted to the apparently single white dwarfs in the sample for which there is {\it Spitzer} photometry at 4.5\,$\upmu$m, and the subset of these 
that have {\it Hubble} data (see Table \ref{tab:TAB2}), in order to determine the frequency of infrared excesses and atmospheric metals that can be attributed to a planetary system.
The nominal frequency of infrared excesses at single stars due to debris discs within the sample is 3/195 (1.5$_{-0.5}^{+1.5}$\,per cent) 
whereas it is 65/143 ($45\pm4$\,per cent) for atmospheric metals, with the uncertainties calculated using the binomial probability 
distribution with a $1\upsigma$ confidence level. As both observations are indicators of circumstellar debris, it is interesting to see that 
{\it the overwhelming majority of stars with debris discs do not exhibit an infrared excess.}

These results broadly agree with previous infrared studies, but at the lower end of estimates made over the years since the launch 
of {\it Spitzer}. Some of the earliest estimates determined infrared excess frequencies due to planetary debris discs of 1--2\,per cent \citep{Mullally,Farihi2009b,
Debes2011}, and this study now robustly establishes this fact. Interestingly, there have been estimates as high as 4\,per cent, even 
including a substantial subset of the stars studied here \citep{Barber,Rocchetto}, and thus the benefit of the largest sample
available, homogeneous selection and data, plus careful vetting are evident. This result holds for the {\em Hubble} subsample 
where the infrared excess frequency at single stars due to planetary debris is 1.4$_{-0.4}^{+1.8}$\,per cent. 

\begin{figure}
\includegraphics[width=\linewidth]{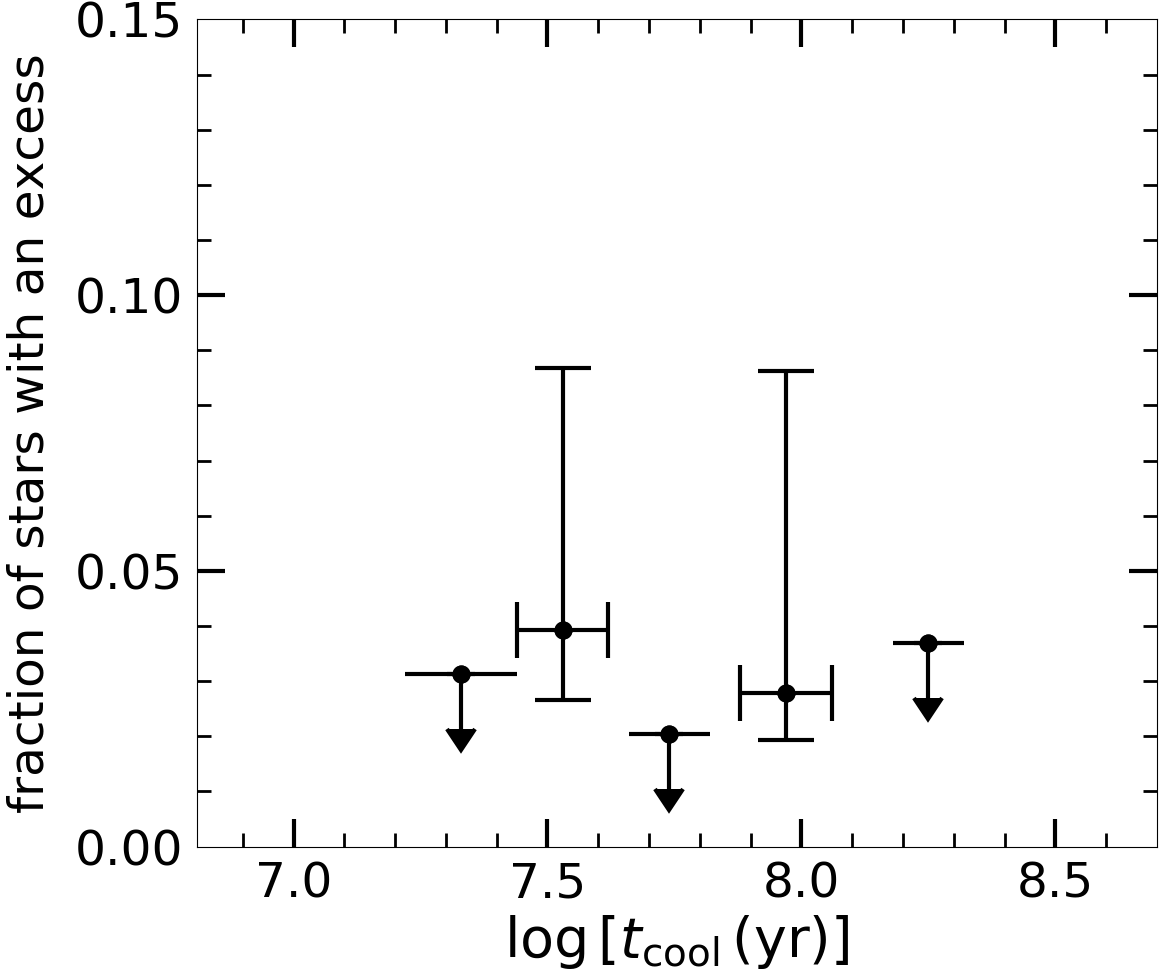}
\caption{The fraction of single stars in the sample with an infrared excess due to a debris disc, with $1\upsigma$ binomial probability errors and upper limits, for several cooling age bins.} 
\label{fig:FIG5}
\end{figure}

The wide temperature range of this unbiased sample is unprecedented, and permits an assessment of the frequency of infrared 
excesses at single stars due to debris discs as a function of post-main sequence planetary system age. Fig.~\ref{fig:FIG5} plots the fraction of single white dwarfs with infrared 
excesses due to debris discs as a function of cooling age. To the level of the available data, the infrared excess frequencies around single stars due to debris discs appear to be flat 
across the entire cooling age range probed in this study, and of the order of a few per cent. 

The combined use of {\em Spitzer} infrared imaging photometry and {\em Hubble} far-ultraviolet spectroscopy is unprecedented for such a large sample of white dwarfs. Also, these two telescopes provide the most sensitive observations of atmospheric metals and infrared excesses, and therefore comparison with other studies will be biased. For the temperatures studied here, the results should be highly robust. However for cooler stars that do not emit sufficiently in the far-ultraviolet, the frequency of infrared excesses and atmospheric metals have a mixture of empirical constraints. In the study of \citet{Zuckerman2003}, for stars cooler and older than the present study with $T_{\rm eff} < 14\,000$\,K, there is only one star
out of 70 that did not have a previously known infrared excess (1.4$_{-0.5}^{+3.1}$\,per cent), but 18 of these were newly found to have metal absorption features (26$_{-4}^{+6}$\,per cent). Infrared excess searches around polluted white dwarfs with $T_{\rm eff} < 10\,000$\,K have resulted in only one or two possibly bona fide detections \citep{Farihi2008,Xu2012,Bergfors2014,Debes2019}.

Despite having observed nearly 200 stars, it can be seen in Fig.~\ref{fig:FIG5} that there are cooling age bins without infrared disc 
detections. This is particularly interesting for two reasons. First, the {\em Hubble} observations, as well as many ground-based
studies, have shown that the fraction of polluted white dwarfs is at least an order of magnitude greater than the fraction of stars with 
an infrared excess due to a circumstellar debris disc. Second, early indications appear to favour a flat distribution of atmospheric metals as a function of cooling age 
\citep{Gansicke2012}, although the full effects of radiative levitation need to be reckoned. The dearth of infrared excesses compared 
to atmospheric metals supports the idea that circumstellar discs orbiting white dwarfs are typically relatively narrow, or in other ways 
tenuous \citep{Farihi2010,Rocchetto,Bonsor2017}. This idea may be supported by the narrow, eccentric ring of debris seen transiting WD\,1145+017 \citep{Redfield,Cauley}.

\subsection{Binary white dwarfs}
\label{sec:SEC4.2}

Recent theoretical work has suggested several, long-term dynamical processes involving a binary companion that may significantly contribute to white dwarf atmospheric pollution \citep{Bonsor2015, Hamers2016, Petrovich, Stephan2017, Smallwood}. While it is clear that a stellar wind from a companion in a short-period orbit may deliver heavy elements onto a white dwarf \citep{Debes2006, Tappert2011}, for systems with sufficiently wide orbits any detectable pollution requires a non-stellar reservoir of material \citep{Farihi2013b,Veras2018}. Auspiciously, white dwarf stars are highly transparent to companion detection via their Earth-sized radii and blue-peaked stellar continua; only the coldest and lowest mass substellar companions remain out of reach of conventional, ground-based detection \citep*{Farihi2005}.  

In the following subsections, the 40 science targets in known or suspected binaries are evaluated. A search using both {\em Gaia} DR2 and via the literature recovers 22 known spatially-resolved and common proper-motion companions (e.g. \citealt{ElBadry2018}), and no new candidates, whereas 18 other white dwarfs are confirmed or suspected to be in short-period binaries. Spatially-unresolved or marginally-resolved IRAC targets (25 sources, including all suspected close binaries) did not produce reliable photometry of the white dwarf itself and are removed from the infrared statistics, leaving 15 white dwarfs in wide binaries that can be searched for photometric excess due to circumstellar dust. However, in the far-ultraviolet with COS, only the white dwarf contributes flux, and thus all 22 sources that are spatially resolved in {\em Gaia} DR2 are retained in the study of pollution frequencies. Below, infrared excesses and atmospheric pollution are discussed sequentially, using the double-blind study together with independent samples of stars to assess the potential relationships of these phenomena with white dwarfs in wide binaries. The resulting statistics are listed in Table~\ref{tab:TAB2}.

\begin{table}
\begin{center}
\caption{Subsample statistics for infrared excess and pollution.\label{tab:TAB2}}
\begin{tabular}{lrrc}

\hline    
Description                         &$n$     &$m$    &$m/n$\\

\hline

\multicolumn{3}{l}{\bf Singles in this study:}\\ 
Infrared excess $\upchi>3$          &195    &3      &0.015\\ 
Metals detected by {\em HST}        &143    &65     &0.455\\ 

\multicolumn{3}{l}{\bf Binaries in this study:}\\ 
Infrared excess $\upchi>3$          &15     &0      &0    \\ 
Metals detected by {\em HST}        &12     &8      &0.667\\ 

\hline
\multicolumn{3}{l}{\bf Published data:}\\ 
Wide companions to dusty stars      &40     &0      &0    \\ 

\hline

\multicolumn{3}{l}{\bf 20\,pc sample:}\\ 
Metals among binary stars           &29     &5      &0.172\\ 
Binaries among polluted stars       &23     &5      &0.217\\ 
Binaries in the 20\,pc stars        &139    &29     &0.209\\ 
\hline

\end{tabular}
\end{center}

\flushleft
{\em Notes.} The second and third columns are the sample size and those targets that meet the condition, respectively,
while the last column is the fraction that satisfy the criterion.

\end{table}

The following evaluates available data on infrared excesses and wide binarity among white dwarfs.  Within the infrared study presented here, all targets confirmed or suspected to be in wide binaries were analysed in an identical manner to the single stars. For the 15 white dwarfs in wide binaries (14 systems; both white dwarfs in 2220+217 are in the sample) that are resolved in their respective IRAC images, there are none with significant flux or colour excess. This yields a binomial probability upper limit of 8\,per cent for infrared excesses due to debris discs around white dwarfs with a wide binary companion. While this result is not as robust as the frequency for single white dwarfs, below a much stronger statement can be made based via confirmed dusty stars with infrared excesses. 

Starting from 40 polluted white dwarfs with well-established infrared excesses (e.g. Table 3 of \citealt{Rocchetto}), {\it Gaia} DR2 enables the straightforward detection of wide companions such as those favoured by dynamical models to enable pollution mechanisms.  This search of known dusty white dwarfs returns no bona fide, co-moving companion candidates.\footnote{The only binary among dusty and polluted white dwarfs is the short-period system SDSS\,J155720.77+091624.6 \citep*{Farihi2017}, where the disc must be {\em circumbinary}, and hence the companion is not suspected of dynamical delivery of the observed material.} [Interestingly and although speculative, a quasi-co-moving companion to SDSS\,J073842.57+183509.6 ($\varpi = 5.8\pm0.2$\,mas, $\upmu = (12.8\pm0.3,-24.1\pm0.2)$\,mas\,yr$^{-1}$) was identified at a projected separation of 0.54\,pc and of approximate mid-M spectral type ($\varpi = 5.9\pm0.1$\,mas, $\upmu = (9.8\pm0.1,-23.7\pm0.1)$\,mas\,yr$^{-1}$). No claim is made here that the pair are related or evaporating, but it is noteworthy that even in the case this system was previously bound, the white dwarf likely evolved essentially as a singleton rather than a binary.] Therefore, it does not appear that binarity has any physical prominence in the generation of dusty debris discs orbiting white dwarfs. 

The possible connection between atmospheric pollution and the presence of a wide companion is explored in the following paragraphs.  In the  double-blind study sample of white dwarfs, eight out of 12 wide binary targets exhibit atmospheric metals in either {\em Hubble} Snapshot or published ultraviolet observations, resulting in a frequency of 67$_{-15}^{+10}$\,per cent. While the nominal value may seem high, it is within $2\upsigma$ of the pollution frequency robustly measured for single stars  \citep{Zuckerman2010,Koester2014}.  Again due to small number statistics of the present study, below a more robust assessment can be made based on larger and well-studied samples.

An independent assessment of metal pollution and wide binarity can be made using the 20\,pc white dwarf sample that is thought to be complete based in {\em Gaia} DR2 \citep{Hollands2018b}.  First, the probability of atmospheric pollution given binarity is evaluated.  There are 29 binary systems containing at least one white dwarf, where five systems have a component with atmospheric metal pollution. Thus, in the 20\,pc sample there is a 17$_{-4}^{+9}$\,per cent probability of pollution among white dwarfs in wide binaries (this is actually a lower limit as not all stars have ultraviolet observations).  It is noteworthy that this frequency is nearly identical to that derived solely on the basis of high-powered optical and ultraviolet spectroscopic surveys \citep{Zuckerman2014}. 

Second, the probability of binarity given atmospheric pollution is calculated.  The 20\,pc sample \citep{Hollands2018b} contains 29 multiples among 139 systems, and thus has a multiplicity fraction of 21$_{-3}^{+4}$\,per cent.  Within the same sample, there are 23 polluted white dwarfs where five are in binaries (22$_{-6}^{+11}$\,per cent). Both these multiplicity fractions are consistent with each other, and with that of deeper field star surveys for a larger number of sources \citep{Farihi2005}. Therefore, it appears that polluted white dwarfs are members of binary systems at the same frequency as field white dwarfs, and that binarity does not play a fundamental role in their nature as metal-enriched stars. 

\subsection{Notes on Individual Objects}
\label{sec:SEC4.3}

\begin{figure}
\includegraphics[width=\linewidth]{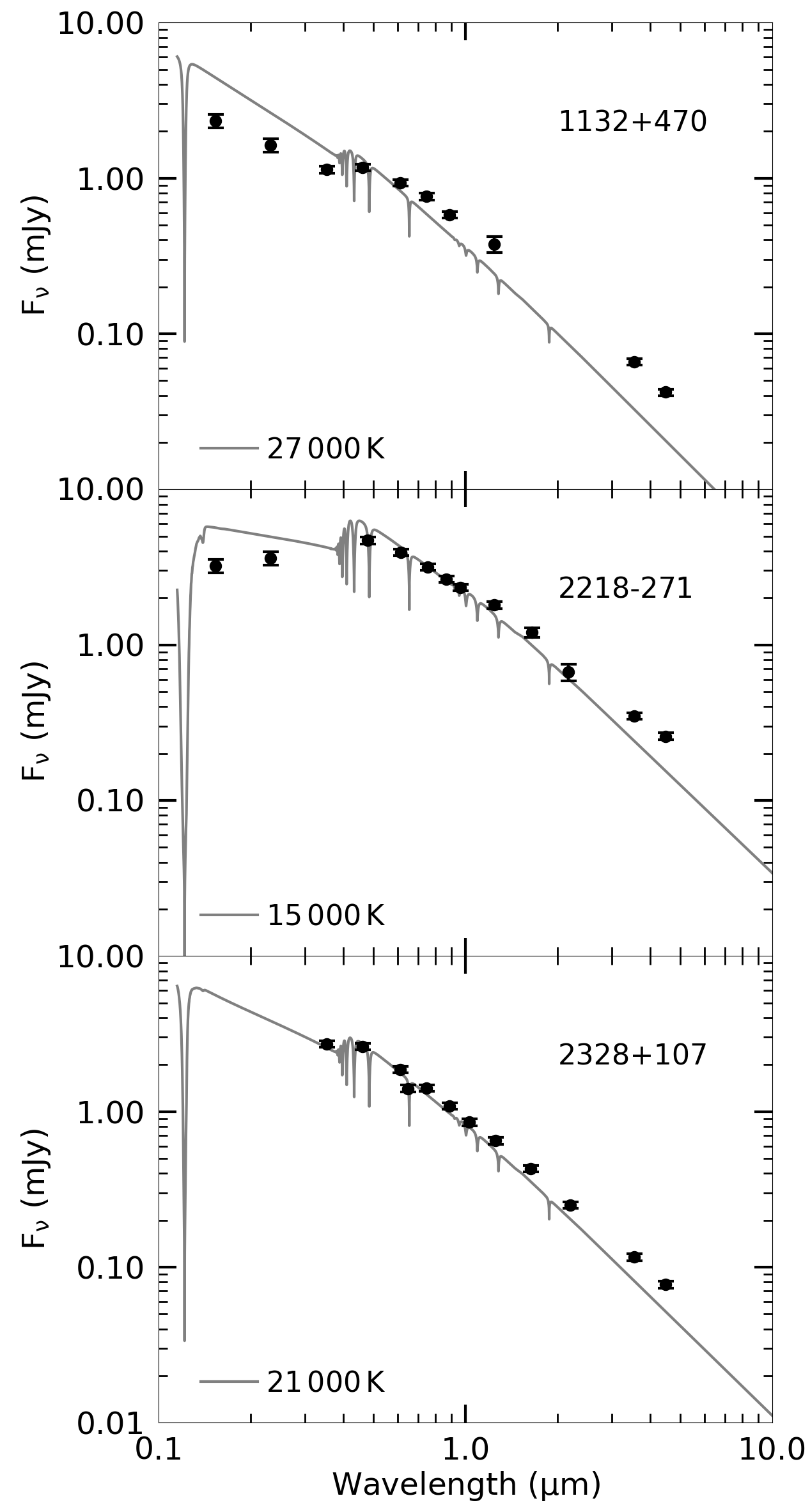}
\caption{SEDs of three stars with significant infrared flux excesses. Data points with error bars are the {\em Spitzer} photometry
together with shorter wavelength fluxes taken from the literature, and the grey lines represent the white dwarf atmospheric models. 
All models have log\,$g = 8$ with $T_{\rm{eff}}$ given in the plots. {\em GALEX} photometry is shown, but excluded from the
fitting process, as observations can suffer from interstellar extinction.}
\label{fig:FIG6}
\end{figure}

Five white dwarfs in the sample have previously identified infrared excesses \citep{Jura2007,Farihi2009b,Xu2012,Rocchetto}. 
Four stars (0843+516, 1015+161, 1018+410, 
and 1457--086) show metal features in COS ultraviolet spectra \citep{Gansicke2012, Koester2014}. One disc candidate exhibits a particularly modest flux excess and is discussed 
below. Finally, 2328+107 displays a subtle ($\upchi>4\upsigma$) flux excess in both warm IRAC channels, however
it {\em does not exhibit atmospheric metals in the ultraviolet}, as obtained with COS.  This star is also discussed below.

{\em 1132+470}.  This star is one of two white dwarfs in the USNO parallax programme that exhibited unambiguous residual astrometric
motion over decades of observation. Both this object and 1639+153 (= LHS\,3236) have clear astrometric periods of several years,
and with companions identified as white dwarfs via adaptive optics \citep{subasavage,harris}. This is consistent with the 
finding that the photometric and spectroscopic effective temperatures are highly discrepant \citep*{Bedard}. The SED shown in
Fig.~\ref{fig:FIG6} demonstrates the spectroscopic $T_{\rm eff}$ is not a good fit to the available multi-wavelength photometry.

{\em 2218--271}.  This target exhibits a significant infrared flux excess (Fig.~\ref{fig:FIG6}), but no photospheric metals are detected 
in ultraviolet spectroscopy with COS. There is no published study that indicates binarity for this star, but there appears to be photometric
excess in both the near-infrared and longer wavelengths. This fact makes it unlikely that the {\em Spitzer} observations suffered from 
background contamination, but this possibility was tested by measuring PSF roundness with the {\sc iraf} task {\sc daofind}. The IRAC 
4.5\,$\upmu$m image of 2218--271 has a roundness value of 0.06, where the average of 20 field stars in the same image is $0.08
\pm0.10$ (a value of 0 is perfectly round). Moreover, the {\it Spitzer} sample of single stars has an average roundness 
of $0.09\pm0.05$, and thus the source of the excess is likely due to an unresolved stellar or substellar companion.

During the ESO SN Ia Progenitor Survey (SPY; \citealt{napiwotzki}), this system was observed twice with UVES, separated by 16\,d, 
in order to probe for short-period radial velocity variations. While the UVES heliocentric-corrected velocities of the H\,$\upalpha$ lines agree 
within the errors, there is a promising 5.3\,h period signal (S.~G.~Parsons, private communication) in photometric data from the 
Catalina Real-time Transient Survey (CRTS; \citealt{Drake}). Further observations are needed to confirm this candidate photometric 
period, and to further constrain the potential binary properties of this system. The {\em Gaia} DR2 astrometric data for 2218--271 are consistent with a single source.

{\em 2328+107}.  The infrared excess toward this star has been previously reported \citep{Rocchetto} and can be seen in 
Fig.~\ref{fig:FIG6}. While the initial discovery was fitted with a relatively narrow disc model, there are no atmospheric metals present 
in the ultraviolet COS spectrum. Similar to 2218--271, this source has no previously published indications of binarity, and exhibits 
photometric excess not only at warm IRAC wavelengths, but also in the near-infrared compared to the adopted atmospheric model. 
An assessment of IRAC 4.5\,$\upmu$m PSF roundness is performed for 2328+107, where a value of 0.02 was determined, which 
agrees well with the value determined for the single white dwarfs observed in the study as well as 20 field stars ($0.09\pm0.04$) within the same 
image. Thus while the observed infrared excess is possibly, but unlikely, due to contamination, it may be more likely that 2328+107 has a stellar or substellar companion.

To assess the possibility of a short-period companion, the radial velocities of H\,$\upalpha$ are compared between two SPY UVES 
spectra taken 3\,d apart. No significant variations are seen in heliocentric-corrected data, and {\em Gaia} DR2 astrometry is 
consistent with that of a single source. Further observations are required to confirm or rule out the possible phase space for 
companions to this white dwarf, as photometric data from CRTS do not reveal any obvious variability.

{\em 1457--086}.  This is one of the most highly metal-polluted white dwarfs known \citep{Koester2005}. However, archival 
2010 and 2013 adaptive optics images from the VLT reveal the presence of a background source within 0.4\,arcsec of the star 
\citep{Dennihy}. This calls into question the nature of the observed infrared excess, and it is unclear to what extend the faint 
background source -- detected so far only in the $J$-band -- contributes to the SED at IRAC wavelengths.

As above, the PSF roundness at 4.5\,$\upmu$m is examined in the 2006 IRAC image for evidence of additional sources. 
The target has a roundness of 0.07 that is comparable to that for 20 field stars in the same image ($0.05\pm0.04$), and for the single star sample. However, there is a second IRAC epoch of this degenerate taken in 2017, and while the 4.5\,$\upmu$m 
flux is unchanged from that measured in 2006, the PSF roundness of the target is 0.54 in 2017. The average roundness for 20 
field stars in the same 2017 image is $0.06\pm0.05$, and hence the elongated PSF is real.  The {\em Gaia} DR2 proper motion 
of 1457--086 is used together with the 2010 and 2013 NACO images to calculate the separation of background source over
time, where it should be 0.1\,arcsec in the 2006 IRAC image, and 0.7\,arcsec in 2017. This is consistent with the roundness values over time.

The chance that the observed 30\,$\upmu$Jy excess at 4.5\,$\upmu$m is caused by a background galaxy within 2.4\,arcsec$^{2}$ 
of the target is one out of 760, following the method outlined in \citet{Farihi2008}. Thus the probability that at least one star in the 
(195-star) sample has a background contamination similar to the excess seen at 1457--086 is 23\,per cent. It should be noted 
that even in the case that 1457--086 has no detectable infrared excess beyond that of the background source, it is clear from the 
ongoing metal pollution that circumstellar material is present. This highlights the overall finding of this study, that discs are likely 
tenuous and subtle, while metal pollution is a much more sensitive tracer of circumstellar material.

\section{Conclusions}
\label{sec:SEC5}

A large and unbiased sample of white dwarfs was observed to determine the frequency of infrared excesses and atmospheric metals using {\it Spitzer} and
{\it Hubble}, respectively. This study focused on the infrared data, where photometric measurements were evaluated using two methods, 
flux excess and colour excess. The flux excess proved more sensitive and seven sources were identified in this manner, while the colour excess 
method only discovered three candidates. Two stars with previously discovered infrared excesses (1457--086 and 2328+107) have a 
flux excess, but not a colour excess, and it is possible the photometric excesses arise from non-disc sources (i.e. companions). 
A further two stars without a colour excess are candidate binaries, as this is consistent with a steep slope at IRAC wavelengths.
If correct, then only 0843+516, 1015+161, and 1018+410 remain as bona fide disc detections in the sample. Interestingly, these three stars are the only sources with greater than $3\upsigma$ excesses via both methods.
Future facilities that are sensitive to fainter infrared disc emission might benefit from a combination of these methods to distinguish genuine circumstellar discs from companions or background sources.

The frequency of infrared excesses due to debris discs over the single white dwarf sample is 1.5$_{-0.5}^{+1.5}$\,per cent, and broadly agrees with previous surveys, but on the lower end of prior estimates. The unbiased nature, size, and sensitivity of the study should make the result fairly definitive and 
robust relative to prior work. Another key result is that, in stark contrast to the modest infrared excess frequency, $45\pm4$\,per cent of single stars in 
the {\it Hubble} COS subsample have atmospheric metals. This confirms prior studies indicating that there is a large population of debris 
discs orbiting white dwarfs, but only a small fraction of these -- roughly one out of 30 -- are detectable in the infrared. The detection of 
atmospheric metals should be considered equivalent to the detection of (prior or current) circumstellar material.

For the white dwarfs in wide binaries, the infrared excess frequency is consistent with zero, but with an upper limit of 8\,per cent. Atmospheric metals occur in 67$_{-15}^{+10}$\,per cent of similar sources, consistent with that measure for single stars in this study.
Notably, the fraction of the local white dwarf population that is both polluted and in a binary is perfectly consistent with the fraction of field white dwarfs in binaries. 
Lastly and remarkably, the known dusty white dwarfs lack co-moving companions and thus their wide binary fraction (0 out of 40; binomial probability upper limit of 2.8\,per cent) is strongly incompatible with that of the field. 
Thus, it does not appear that binarity is connected -- even weakly -- with atmospheric pollution in white dwarf stars. 
On the basis of this study and these facts, the delivery of planetary debris into the immediate vicinity of white dwarf stars and the subsequent surface pollution is likely controlled by major planets.

\section*{Acknowledgements}

The authors would like to thank the anonymous referee for helpful comments that improved the manuscript. This work is based on observations made with the {\em Spitzer Space Telescope}, which is operated by the Jet Propulsion Laboratory, 
California Institute of Technology under a contract with NASA. Based on observations made with the NASA/ESA {\em Hubble Space 
Telescope}, obtained at the Space Telescope Science Institute, which is operated by the Association of Universities for Research in 
Astronomy, Inc., under NASA contract NAS 5-26555. These observations are associated with programs \#12169, \#12474, \#13652, 
and \#14077. T.~G.~Wilson and A.~Swan acknowledge funding from STFC studentships. B.~T.~G{\"a}nsicke received financial support 
from the European Union's European Research Council, grant number 320964 (WDTracer).




\bibliographystyle{mnras}
\bibliography{IRACCOS2} 




\appendix
\section{Supplementary Material}

\onecolumn
\label{sec:A1}

\begin{ThreePartTable}
\begin{TableNotes}
\item\textit{Notes.} Cooling ages are calculated using white dwarf evolutionary models \citep*{Fontaine2001} with stellar parameters derived from {\em Gaia} DR2 \citep{GentileFusillo2019}. Fluxes without reported errors are upper limits. Confirmed and candidate companions are labelled as MS (= main-sequence star, most of which are M dwarfs), and WD (= white dwarf).
\item$^{\rm a}$ Common proper-motion binary resolved in IRAC images. Reliable photometry was recovered for the target via aperture photometry or {\sc apex}.
\item$^{\rm b}$ Common proper-motion binary resolved in {\it Gaia} DR2. IRAC images fail to reveal both sources, and thus the point-source photometry reported may not be reliable. 
\item$^{\rm c}$ Confirmed or candidate short-period binary. {\it Gaia} DR2 does not resolve any additional components, and IRAC images are point-like. 
\item$^{\rm d}$ Photometry contaminated by background source. 
\item\textbf{References.} (1) \cite{Zuckerman1992}; (2) \cite{Koester2009a}; (3) \cite{Farihi2005}; (4) \cite{Maoz}; (5) \cite{Mueller}; (6) \cite{ElBadry2018}; (7) \cite*{Maxted}; (8) \cite{Jordan1998}; (9) \cite{Wachter}; (10) \cite{Schultz}; (11) \cite{Greenstein1984}; (12) \cite{Eggen1965a}; (13) \cite*{Ferguson}; (14) \cite{Cheselka}; (15) \cite{Karl}; (16) \cite{Hoard2007}; (17) \cite{Greenstein1976}; (18) \cite{Eggen1965b}; (19) \cite*{Kidder}; (20) \cite{Greenstein1974}.
\end{TableNotes}

\begin{longtable}{cccccccccc}
  \caption{{\it Spitzer} IRAC fluxes for white dwarfs in this sample, together with an indication of atmospheric Si\,{\sc ii} observed by {\it Hubble} COS. Stellar parameters and binary type are taken from the literature.}\\ 

  \hline              
  WD & $T_{\rm{eff}}$ & $m$ & $\rm{log}$ & $F_{\rm{3.6\,\upmu m}}$ & $F_{\rm{4.5\,\upmu m}}$ & $F_{\rm{5.7\,\upmu m}}$ & $F_{\rm{7.9\,\upmu m}}$ & Si\,{\sc ii} & Binarity \\
  & (K) & mag) & [$t_{\rm{cool}}\,$(yr)] & ($\upmu$Jy) & ($\upmu$Jy) & ($\upmu$Jy) & ($\upmu$Jy) &  &  \\
  \hline\hline
  \endfirsthead
  
  \multicolumn{6}{l}{{\bfseries \tablename\ \thetable{} -- continued from previous page}} \\
  \hline                  
  WD & $T_{\rm{eff}}$ & $m$ & $\rm{log}$ & $F_{\rm{3.6\,\upmu m}}$ & $F_{\rm{4.5\,\upmu m}}$ & $F_{\rm{5.7\,\upmu m}}$ & $F_{\rm{7.9\,\upmu m}}$ & Si\,{\sc ii} & Binarity \\
  & (K) & (mag) & [$t_{\rm{cool}}\,$(yr)] & ($\upmu$Jy) & ($\upmu$Jy) & ($\upmu$Jy) & ($\upmu$Jy) &  & \\
  \hline\hline
  \endhead

  \hline
  \insertTableNotes
  \endlastfoot
  
  0000+171 & $20\,200$ & $15.8$ & $7.76$ & $64\pm3$ & $44\pm2$ & $-$ & $-$ & Y & $-$ \\
  0002+165 & $25\,900$ & $15.7$ & $7.28$ & $65\pm3$ & $40\pm2$ & $-$ & $-$ & Y & $-$ \\
  0004+061 & $24\,400$ & $16.1$ & $7.91$ & $55\pm3$ & $33\pm2$ & $-$ & $-$ & N & $-$ \\
  0013--241 & $18\,500$ & $15.4$ & $8.00$ & $103\pm5$ & $66\pm4$ & $-$ & $-$ & N & $-$ \\
  0017+061$^{\rm b}$ & $28\,100$ & $15.2$ & $7.08$ & $2400\pm100$ & $1620\pm80$ & $-$ & $-$ & $-$ & MS, 1 \\
  0018--339 & $20\,600$ & $14.6$ & $7.69$ & $200\pm10$ & $120\pm6$ & $-$ & $-$ & N & $-$\\
  0028--474$^{\rm c}$ & $17\,400$ & $15.2$ & $7.87$ & $166\pm8$ & $104\pm5$ & $-$ & $-$ & N & WD, 2 \\
  0047--524 & $18\,800$ & $14.2$ & $7.83$ & $-$ & $190\pm10$ & $-$ & $66\pm3$ & N & $-$\\
  0048--544 & $17\,900$ & $15.2$ & $7.87$ & $125\pm6$ & $81\pm4$ & $-$ & $-$ & $-$ & $-$ \\
  0048+202 & $20\,400$ & $15.4$ & $7.83$ & $102\pm5$ & $65\pm3$ & $-$ & $-$ & $-$ & $-$  \\
  0052--147 & $26\,700$ & $15.1$ & $7.68$ & $106\pm5$ & $68\pm3$ & $-$ & $-$ & N & $-$ \\
  0059+257 & $21\,400$ & $15.9$ & $7.77$ & $68\pm4$ & $44\pm2$ & $-$ & $-$ & Y & $-$ \\
  0102+095 & $24\,800$ & $14.4$ & $7.45$ & $220\pm10$ & $147\pm7$ & $-$ & $-$ & Y & $-$ \\
  0106--358$^{\rm a}$ & $30\,900$ & $14.7$ & $7.04$ & $159\pm8$ & $91\pm5$ & $-$ & $-$ & Y & MS, 3 \\
  0110--139 & $26\,300$ & $15.8$ & $7.38$ & $60\pm3$ & $40\pm2$ & $-$ & $-$ & $-$ & $-$ \\ 
  0114--605$^{\rm c}$ & $24\,700$ & $15.1$ & $7.30$ & $113\pm6$ & $75\pm4$ & $-$ & $-$ & Y & WD, 4 \\
  0124--257 & $23\,000$ & $16.2$ & $7.43$ & $36\pm2$ & $26\pm1$ & $-$ & $-$ & N & $-$ \\
  0127--050 & $16\,800$ & $14.9$ & $8.16$ & $177\pm9$ & $112\pm6$ & $-$ & $-$ & $-$ & $-$ \\
  0127+270 & $24\,900$ & $15.9$ & $7.36$ & $55\pm3$ & $33\pm2$ & $25$ & $38$ & $-$ & $-$ \\
  0129--205 & $20\,000$ & $15.3$ & $7.79$ & $104\pm5$ & $68\pm4$ & $-$ & $-$ & $-$ & $-$ \\
  0131+018$^{\rm c}$ & $15\,200$ & $14.5$ & $8.40$ & $340\pm20$ & $220\pm10$ & $-$ & $-$ & N & WD, 2 \\
  0136+768 & $16\,900$ & $14.8$ & $7.95$ & $230\pm10$ & $151\pm8$ & $-$ & $-$ & N & $-$ \\
  0140--392 & $21\,800$ & $14.4$ & $7.62$ & $250\pm10$ & $165\pm8$ & $-$ & $-$ & Y & $-$ \\
  0145--257$^{\rm b}$ & $26\,700$ & $14.5$ & $7.20$ & $8100\pm400$ & $5400\pm300$ & $-$ & $-$ & $-$ & MS, 5 \\
  0155+069 & $22\,000$ & $15.5$ & $7.41$ & $108\pm6$ & $68\pm3$ & $-$ & $-$ & N & $-$ \\
  0200+248 & $23\,300$ & $15.7$ & $7.79$ & $110\pm10$ & $71\pm8$ & $-$ & $-$ & N & $-$ \\
  0201--052 & $24\,600$ & $16.0$ & $7.32$ & $41\pm2$ & $28\pm2$ & $-$ & $-$ & $-$ & $-$ \\
  0216+143 & $26\,900$ & $14.6$ & $7.26$ & $210\pm10$ & $130\pm7$ & $80\pm10$ & $33$ & N & $-$ \\
  0221--055$^{\rm b}$ & $25\,800$ & $16.2$ & $7.52$ & $62\pm3$ & $42\pm2$ & $-$ & $-$ & $-$ & WD, 4 \\
  0222--265 & $23\,200$ & $15.7$ & $7.60$ & $76\pm4$ & $51\pm3$ & $-$ & $-$ & $-$ & $-$ \\
  0227+050 & $19\,300$ & $12.8$ & $7.83$ & $-$ & $780\pm40$ & $-$ & $260\pm10$ & $-$ & $-$ \\
  0229+270 & $24\,200$ & $15.5$ & $7.36$ & $80\pm4$ & $51\pm3$ & $-$ & $-$ & $-$ & $-$ \\
  0231--054 & $17\,300$ & $14.3$ & $8.86$ & $-$ & $240\pm10$ & $-$ & $71\pm4$ & N & $-$ \\
  0242--174 & $20\,700$ & $15.4$ & $7.79$ & $96\pm5$ & $62\pm3$ & $-$ & $-$ & Y & $-$ \\
  0300--232 & $22\,400$ & $15.7$ & $8.03$ & $72\pm4$ & $44\pm2$ & $-$ & $-$ & $-$ & $-$ \\
  0305--117 & $26\,800$ & $16.0$ & $7.18$ & $53\pm3$ & $33\pm2$ & $-$ & $-$ & N & $-$ \\
  0307+149$^{\rm a}$ & $21\,400$ & $15.4$ & $7.61$ & $112\pm6$ & $71\pm4$ & $-$ & $-$ & Y & WD, 6 \\
  0308--230 & $23\,600$ & $15.1$ & $8.14$ & $108\pm5$ & $70\pm4$ & $-$ & $-$ & N & $-$ \\
  0308+188 & $18\,500$ & $14.2$ & $7.89$ & $320\pm20$ & $200\pm10$ & $-$ & $-$ & N & $-$ \\
  0316+345 & $15\,400$ & $14.2$ & $7.98$ & $-$ & $230\pm10$ & $-$ & $74\pm9$ & N & $-$ \\
  0331+226 & $21\,500$ & $15.3$ & $7.43$ & $110\pm6$ & $71\pm4$ & $-$ & $-$ & $-$ & $-$ \\
  0341+021$^{\rm c}$ & $22\,200$ & $15.4$ & $7.27$ & $102\pm5$ & $61\pm3$ & $-$ & $-$ & Y & WD, 7 \\
  0345+134 & $25\,100$ & $16.1$ & $7.56$ & $46\pm2$ & $29\pm2$ & $-$ & $-$ & $-$ & $-$ \\
  0349--256 & $21\,000$ & $15.7$ & $7.73$ & $72\pm4$ & $47\pm2$ & $-$ & $-$ & $-$ & $-$ \\
  0352+018 & $22\,100$ & $15.6$ & $7.28$ & $79\pm4$ & $50\pm3$ & $-$ & $-$ & Y & $-$ \\
  0358--514 & $23\,400$ & $15.7$ & $7.46$ & $70\pm4$ & $47\pm2$ & $-$ & $-$ & N & $-$ \\
  0400+148 & $14\,600$ & $14.9$ & $8.54$ & $200\pm10$ & $130\pm7$ & $-$ & $-$ & N & $-$ \\
  0403--414 & $22\,700$ & $16.4$ & $7.54$ & $41\pm2$ & $26\pm1$ & $-$ & $-$ & Y & $-$ \\
  0406+169 & $15\,800$ & $15.3$ & $8.46$ & $126\pm6$ & $78\pm4$ & $53\pm5$ & $44\pm6$ & N & $-$ \\
  0410+117$^{\rm a}$ & $21\,100$ & $13.9$ & $7.58$ & $-$ & $250\pm10$ & $-$ & $64\pm3$ & Y & MS, 6 \\
  0414--406 & $20\,900$ & $16.1$ & $7.79$ & $48\pm3$ & $30\pm2$ & $-$ & $-$ & N & $-$ \\
  0416--105 & $24\,900$ & $15.4$ & $7.26$ & $87\pm4$ & $53\pm3$ & $-$ & $-$ & Y & $-$ \\
  0418--534 & $27\,100$ & $16.4$ & $7.15$ & $32\pm2$ & $21\pm1$ & $-$ & $-$ & Y & $-$ \\
  0418--103 & $23\,400$ & $15.7$ & $7.94$ & $70\pm4$ & $44\pm2$ & $-$ & $-$ & N & $-$ \\
  0421+162 & $19\,600$ & $14.3$ & $7.93$ & $280\pm10$ & $172\pm9$ & $115\pm6$ & $60\pm5$ & Y & $-$ \\
  0431+126 & $21\,400$ & $14.2$ & $7.75$ & $-$ & $177\pm9$ & $-$ & $34\pm7$ & Y & $-$ \\
  0452--347 & $21\,200$ & $16.1$ & $7.62$ & $57\pm3$ & $37\pm2$ & $-$ & $-$ & Y & $-$ \\
  0455--532 & $24\,400$ & $16.8$ & $7.22$ & $31\pm2$ & $18\pm1$ & $-$ & $-$ & $-$ & $-$ \\
  0507+045A$^{\rm a}$ & $20\,800$ & $14.2$ & $7.82$ & $-$ & $210\pm10$ & $-$ & $140\pm10$ & Y & WD, 8 \\
  0730+487 & $14\,900$ & $14.8$ & $8.78$ & $210\pm10$ & $129\pm7$ & $86\pm6$ & $56$ & N & $-$ \\
  0732--427 & $15\,600$ & $14.2$ & $8.39$ & $-$ & $250\pm10$ & $-$ & $90\pm20$ & N & $-$ \\
  0816+297 & $16\,700$ & $15.8$ & $8.06$ & $78\pm4$ & $52\pm3$ & $26$ & $26$ & N & $-$ \\
  0817+386 & $25\,200$ & $15.8$ & $7.30$ & $65\pm4$ & $42\pm2$ & $-$ & $-$ & N & $-$ \\
  0821+632 & $16\,800$ & $15.9$ & $8.01$ & $72\pm4$ & $46\pm2$ & $-$ & $-$ & N & $-$ \\
  0839+231 & $25\,800$ & $14.5$ & $7.20$ & $210\pm10$ & $131\pm7$ & $-$ & $-$ & Y & $-$ \\
  0843+516 & $23\,900$ & $16.0$ & $7.56$ & $136\pm7$ & $137\pm7$ & $102\pm7$ & $159\pm9$ & Y & $-$ \\
  0846+557 & $27\,400$ & $16.4$ & $7.11$ & $33\pm2$ & $18\pm1$ & $-$ & $-$ & N & $-$ \\
  0854+404 & $22\,300$ & $14.8$ & $7.58$ & $156\pm8$ & $97\pm5$ & $-$ & $-$ & Y & $-$ \\
  0859--039 & $23\,700$ & $13.2$ & $7.38$ & $700\pm40$ & $440\pm20$ & $-$ & $-$ & $-$ & $-$ \\
  0859+337 & $25\,400$ & $16.6$ & $7.49$ & $32\pm2$ & $19\pm1$ & $-$ & $-$ & $-$ & $-$ \\
  0904+391 & $26\,200$ & $16.3$ & $7.40$ & $33\pm3$ & $23\pm2$ & $-$ & $-$ & N & $-$ \\
  0915+526 & $15\,600$ & $15.5$ & $8.27$ & $107\pm5$ & $65\pm3$ & $-$ & $-$ & N & $-$ \\
  0920+363 & $24\,100$ & $16.1$ & $7.22$ & $56\pm3$ & $35\pm2$ & $-$ & $-$ & Y & $-$ \\
  0922+183 & $24\,700$ & $16.4$ & $7.48$ & $40\pm2$ & $24\pm2$ & $-$ & $-$ & $-$ & $-$ \\
  0933+025$^{\rm b}$ & $22\,400$ & $15.9$ & $7.78$ & $3600\pm200$ & $2500\pm100$ & $-$ & $-$ & N & MS, 10 \\
  0938+550 & $18\,500$ & $14.8$ & $8.04$ & $200\pm10$ & $123\pm6$ & $-$ & $-$ & $-$ & $-$ \\
  0944+192 & $17\,400$ & $14.5$ & $8.16$ & $250\pm10$ & $159\pm8$ & $-$ & $-$ & N & $-$ \\
  0947+325 & $22\,100$ & $15.5$ & $8.04$ & $87\pm4$ & $57\pm3$ & $-$ & $-$ & N & $-$ \\
  0954+697 & $21\,400$ & $16.0$ & $7.60$ & $63\pm3$ & $44\pm2$ & $-$ & $-$ & Y & $-$ \\
  0956+020 & $15\,700$ & $15.7$ & $8.13$ & $88\pm5$ & $54\pm3$ & $-$ & $-$ & N & $-$ \\
  1003--023 & $20\,600$ & $15.3$ & $7.83$ & $121\pm6$ & $73\pm4$ & $-$ & $-$ & $-$ & $-$ \\
  1005+642 & $19\,700$ & $13.7$ & $7.88$ & $460\pm20$ & $290\pm10$ & $-$ & $-$ & N & $-$ \\
  1012--008 & $23\,200$ & $15.6$ & $7.72$ & $80\pm4$ & $49\pm3$ & $-$ & $-$ & $-$ & $-$ \\
  1013+256 & $22\,000$ & $16.3$ & $7.68$ & $42\pm2$ & $23\pm1$ & $-$ & $-$ & Y & $-$ \\
  1015+161 & $20\,000$ & $15.6$ & $7.99$ & $200\pm10$ & $167\pm8$ & $142\pm8$ & $124\pm7$ & Y & $-$ \\
  1016--308 & $16\,300$ & $14.6$ & $8.37$ & $240\pm10$ & $153\pm8$ & $-$ & $-$ & $-$ & $-$ \\
  1017+125$^{\rm a}$ & $21\,400$ & $15.7$ & $7.73$ & $73\pm4$ & $49\pm3$ & $28\pm3$ & $19\pm6$ & Y & WD, 3 \\
  1018+410 & $23\,700$ & $16.4$ & $7.82$ & $86\pm4$ & $76\pm4$ & $-$ & $-$ & $-$ & $-$ \\
  1020--207 & $19\,900$ & $15.0$ & $7.78$ & $130\pm7$ & $84\pm4$ & $-$ & $-$ & Y & $-$  \\
  1034+492 & $20\,700$ & $15.4$ & $7.97$ & $101\pm5$ & $59\pm3$ & $49\pm4$ & $30$ & Y & $-$ \\
  1038+633 & $24\,500$ & $15.2$ & $8.00$ & $109\pm5$ & $67\pm3$ & $57\pm4$ & $27$ & Y & $-$ \\
  1049--158 & $20\,600$ & $14.4$ & $8.24$ & $260\pm10$ & $155\pm8$ & $-$ & $-$ & N & $-$ \\
  1049+103$^{\rm c}$ & $20\,600$ & $15.8$ & $7.71$ & $3700\pm200$ & $2600\pm100$ & $-$ & $-$ & N & MS, 1 \\
  1052+273 & $23\,100$ & $14.1$ & $8.04$ & $300\pm10$ & $190\pm10$ & $130\pm7$ & $59\pm7$ & N & $-$ \\
  1058--129 & $24\,300$ & $14.9$ & $8.20$ & $128\pm7$ & $82\pm4$ & $42\pm6$ & $30\pm8$ & N & $-$ \\
  1102+748 & $19\,700$ & $15.1$ & $8.24$ & $136\pm7$ & $90\pm5$ & $-$ & $-$ & N & $-$ \\
  1104+602 & $17\,900$ & $13.7$ & $8.11$ & $460\pm20$ & $290\pm10$ & $-$ & $-$ & N & $-$ \\
  1105--048$^{\rm a}$ & $15\,100$ & $13.1$ & $8.18$ & $-$ & $640\pm30$ & $-$ & $260\pm20$ & $-$ & MS, 11 \\
  1113+413 & $25\,400$ & $15.4$ & $7.30$ & $96\pm5$ & $61\pm3$ & $-$ & $-$ & Y & $-$ \\
  1115+166 & $22\,100$ & $15.1$ & $7.72$ & $134\pm7$ & $86\pm4$ & $65\pm6$ & $44$ & N & $-$ \\
  1117--023 & $14\,700$ & $14.5$ & $8.31$ & $290\pm10$ & $180\pm9$ & $-$ & $-$ & N & $-$ \\
  1120+439 & $27\,000$ & $15.4$ & $7.63$ & $84\pm4$ & $48\pm3$ & $55$ & $50$ & $-$ & $-$ \\
  1122--324 & $21\,700$ & $15.8$ & $7.58$ & $62\pm3$ & $40\pm2$ & $-$ & $-$ & $-$ & $-$ \\
  1126+384 & $25\,200$ & $14.9$ & $7.34$ & $127\pm6$ & $76\pm4$ & $-$ & $-$ & Y & $-$ \\
  1128+564 & $26\,600$ & $16.5$ & $7.18$ & $30\pm2$ & $19\pm1$ & $-$ & $-$ & N & $-$ \\
  1129+155 & $17\,700$ & $14.1$ & $7.98$ & $380\pm20$ & $230\pm10$ & $159\pm9$ & $78\pm7$ & Y & $-$ \\
  1132+470 & $27\,500$ & $16.4$ & $8.28$ & $66\pm3$ & $42\pm2$ & $-$ & $-$ & $-$ & $-$ \\
  1133+293 & $23\,000$ & $14.9$ & $7.52$ & $147\pm7$ & $91\pm8$ & $64\pm7$ & $44\pm8$ & Y & $-$ \\
  1134+300 & $21\,300$ & $12.5$ & $8.22$ & $1380\pm70$ & $890\pm40$ & $-$ & $-$ & $-$ & $-$ \\
  1136+139 & $23\,900$ & $16.8$ & $7.43$ & $23\pm3$ & $16\pm2$ & $-$ & $-$ & $-$ & $-$ \\
  1143+321$^{\rm a}$ & $15\,900$ & $13.7$ & $8.25$ & $740\pm40$ & $480\pm30$ & $-$ & $-$ & N & MS, 12 \\
  1145+187 & $26\,600$ & $14.2$ & $7.18$ & $240\pm10$ & $143\pm7$ & $-$ & $-$ & Y & $-$ \\
  1152+371 & $27\,400$ & $14.7$ & $8.47$ & $13\pm2$ & $10\pm1$ & $-$ & $-$ & $-$ & $-$ \\
  1201--001 & $19\,800$ & $15.2$ & $8.17$ & $124\pm6$ & $80\pm4$ & $48\pm3$ & $18\pm4$ & $-$ & $-$ \\
  1202+308 & $28\,900$ & $16.3$ & $7.04$ & $40\pm2$ & $26\pm2$ & $-$ & $-$ & N & $-$ \\
  1204--322 & $21\,300$ & $15.6$ & $7.79$ & $80\pm4$ & $52\pm3$ & $-$ & $-$ & $-$ & $-$ \\
  1220+234 & $26\,500$ & $15.6$ & $7.74$ & $67\pm3$ & $43\pm2$ & $-$ & $-$ & N & $-$ \\
  1224+309$^{\rm c}$ & $28\,800$ & $16.2$ & $7.08$ & $610\pm30$ & $410\pm20$ & $-$ & $-$ & Y & MS, 13 \\
  1229--012 & $19\,400$ & $14.5$ & $7.60$ & $240\pm10$ & $153\pm8$ & $91\pm6$ & $64\pm8$ & N & $-$ \\
  1230--308 & $22\,800$ & $15.7$ & $7.41$ & $70\pm4$ & $46\pm2$ & $-$ & $-$ & N & $-$ \\
  1232+479 & $14\,400$ & $14.5$ & $8.31$ & $280\pm10$ & $173\pm9$ & $-$ & $-$ & $-$ & $-$ \\
  1233--164$^{\rm c}$ & $24\,900$ & $15.1$ & $7.66$ & $114\pm6$ & $73\pm4$ & $-$ & $-$ & N & WD, 4 \\
  1241+235 & $26\,700$ & $15.7$ & $7.20$ & $76\pm4$ & $53\pm3$ & $-$ & $-$ & $-$ & $-$ \\
  1243+015 & $21\,600$ & $16.5$ & $7.72$ & $34\pm2$ & $23\pm1$ & $-$ & $-$ & Y & $-$ \\
  1247--115 & $28\,100$ & $15.3$ & $7.23$ & $88\pm4$ & $54\pm3$ & $-$ & $-$ & Y & $-$ \\
  1249+160 & $25\,600$ & $14.7$ & $7.24$ & $177\pm9$ & $107\pm5$ & $-$ & $-$ & Y & $-$ \\
  1249+182 & $19\,900$ & $15.2$ & $7.74$ & $95\pm5$ & $60\pm3$ & $-$ & $-$ & Y & $-$ \\
  1257+047$^{\rm a}$ & $21\,800$ & $14.9$ & $7.61$ & $151\pm8$ & $98\pm5$ & $-$ & $-$ & Y & MS, 3 \\
  1305+018$^{\rm b}$ & $29\,000$ & $15.2$ & $7.26$ & $110\pm6$ & $77\pm4$ & $-$ & $-$ & $-$ & MS, 14 \\
  1308--301 & $15\,300$ & $15.2$ & $8.34$ & $155\pm8$ & $96\pm5$ & $-$ & $-$ & Y & $-$ \\
  1310--305 & $20\,400$ & $14.5$ & $7.83$ & $220\pm10$ & $143\pm7$ & $-$ & $-$ & Y & $-$ \\
  1314--153 & $15\,700$ & $14.9$ & $8.16$ & $210\pm10$ & $134\pm7$ & $-$ & $-$ & N & $-$ \\
  1323--514 & $19\,400$ & $14.4$ & $7.80$ & $280\pm10$ & $176\pm9$ & $-$ & $-$ & Y & $-$ \\
  1325--089 & $17\,000$ & $15.0$ & $8.01$ & $165\pm8$ & $107\pm5$ & $-$ & $-$ & Y & $-$ \\
  1325+279 & $21\,300$ & $15.8$ & $7.91$ & $64\pm7$ & $42\pm5$ & $-$ & $-$ & Y & $-$ \\
  1330+473 & $22\,500$ & $15.2$ & $7.57$ & $106\pm6$ & $70\pm4$ & $-$ & $-$ & N & $-$ \\
  1333+497 & $29\,300$ & $15.7$ & $7.08$ & $58\pm3$ & $36\pm2$ & $-$ & $-$ & $-$ & $-$ \\
  1334--160$^{\rm a}$ & $18\,700$ & $15.4$ & $8.27$ & $116\pm7$ & $94\pm6$ & $-$ & $-$ & $-$ & MS, 6 \\
  1334+070$^{\rm c}$ & $16\,900$ & $15.4$ & $7.63$ & $162\pm8$ & $99\pm5$ & $-$ & $-$ & N & WD, 2 \\
  1335+369 & $20\,500$ & $14.8$ & $7.62$ & $210\pm10$ & $135\pm7$ & $-$ & $-$ & $-$ & $-$ \\
  1337+705 & $20\,500$ & $12.6$ & $7.76$ & $-$ & $710\pm40$ & $-$ & $250\pm10$ & $-$ & $-$ \\
  1338+081 & $24\,400$ & $16.4$ & $7.34$ & $31\pm2$ & $19\pm1$ & $-$ & $-$ & $-$ & $-$ \\
  1339+346$^{\rm d}$ & $16\,000$ & $15.9$ & $8.08$ & $1030$ & $-$ & $-$ & $-$ & N & $-$ \\
  1349+144$^{\rm c}$ & $16\,600$ & $15.3$ & $8.16$ & $134\pm7$ & $84\pm4$ & $-$ & $-$ & N & WD, 15 \\
  1353+409 & $23\,500$ & $15.5$ & $7.38$ & $77\pm4$ & $50\pm3$ & $33\pm5$ & $26\pm5$ & N & $-$ \\
  1408+323 & $18\,200$ & $14.0$ & $7.97$ & $-$ & $250\pm10$ & $-$ & $92\pm7$ & Y & $-$ \\
  1412--109$^{\rm b}$ & $25\,700$ & $15.9$ & $7.38$ & $760\pm40$ & $530\pm30$ & $330\pm20$ & $220\pm20$ & Y & MS, 16 \\
  1421+318 & $27\,200$ & $15.4$ & $7.23$ & $82\pm4$ & $52\pm3$ & $-$ & $-$ & Y & $-$ \\
  1433+538$^{\rm c}$ & $22\,400$ & $16.1$ & $7.43$ & $900\pm50$ & $600\pm30$ & $-$ & $-$ & $-$ & MS, 17 \\
  1449+168 & $22\,400$ & $15.4$ & $7.58$ & $91\pm5$ & $58\pm3$ & $-$ & $-$ & $-$ & $-$ \\
  1451+006 & $25\,500$ & $15.3$ & $7.30$ & $99\pm5$ & $63\pm3$ & $-$ & $-$ & Y & $-$ \\
  1452--042 & $23\,500$ & $16.3$ & $7.83$ & $51\pm3$ & $36\pm2$ & $-$ & $-$ & $-$ & $-$ \\
  1452+553 & $28\,300$ & $16.1$ & $7.60$ & $43\pm2$ & $28\pm2$ & $-$ & $-$ & Y & $-$ \\
  1457--086 & $20\,400$ & $15.8$ & $7.72$ & $114\pm6$ & $74\pm4$ & $49\pm6$ & $53$ & Y & $-$ \\
  1459+347 & $21\,500$ & $15.8$ & $8.20$ & $66\pm3$ & $41\pm2$ & $-$ & $-$ & N & $-$ \\
  1507+220 & $19\,900$ & $15.0$ & $7.80$ & $151\pm8$ & $97\pm5$ & $-$ & $-$ & $-$ & $-$ \\
  1508+548 & $17\,000$ & $15.7$ & $8.13$ & $80\pm4$ & $50\pm3$ & $-$ & $-$ & N & $-$ \\
  1511+009 & $27\,600$ & $15.9$ & $7.26$ & $56\pm3$ & $34\pm2$ & $-$ & $-$ & $-$ & $-$ \\
  1513+442 & $29\,200$ & $15.2$ & $7.15$ & $82\pm4$ & $52\pm3$ & $-$ & $-$ & Y & $-$ \\
  1517+373 & $25\,400$ & $16.3$ & $7.38$ & $41\pm2$ & $29\pm2$ & $-$ & $-$ & $-$ & $-$ \\
  1518--003 & $15\,400$ & $15.2$ & $8.16$ & $122\pm6$ & $78\pm4$ & $-$ & $-$ & N & $-$ \\
  1524--749 & $23\,100$ & $16.0$ & $7.36$ & $64\pm3$ & $39\pm2$ & $-$ & $-$ & N & $-$ \\
  1525+257 & $22\,300$ & $15.7$ & $7.97$ & $70\pm4$ & $45\pm2$ & $-$ & $-$ & N & $-$ \\
  1527+090 & $21\,200$ & $14.3$ & $7.72$ & $260\pm10$ & $165\pm8$ & $-$ & $-$ & Y & $-$ \\
  1531--022 & $18\,600$ & $14.0$ & $7.92$ & $-$ & $240\pm10$ & $-$ & $103\pm8$ & N & $-$ \\
  1533--057 & $20\,000$ & $15.4$ & $8.07$ & $95\pm5$ & $58\pm3$ & $-$ & $-$ & Y & $-$ \\
  1535+293 & $24\,500$ & $15.9$ & $7.40$ & $51\pm3$ & $30\pm2$ & $-$ & $-$ & N & $-$ \\
  1539+530$^{\rm b}$ & $26\,800$ & $15.6$ & $7.43$ & $5200\pm300$ & $3400\pm200$ & $-$ & $-$ & $-$ & MS, 3 \\
  1547+057 & $24\,400$ & $15.9$ & $8.00$ & $57\pm3$ & $36\pm2$ & $-$ & $-$ & N & $-$ \\
  1548+149 & $21\,500$ & $15.2$ & $7.74$ & $123\pm6$ & $77\pm4$ & $-$ & $-$ & Y & $-$ \\
  1550+183 & $14\,300$ & $14.8$ & $8.62$ & $220\pm10$ & $136\pm7$ & $95\pm7$ & $53\pm6$ & N & $-$ \\
  1553+353 & $25\,600$ & $14.8$ & $7.30$ & $169\pm8$ & $105\pm5$ & $62\pm5$ & $43$ & $-$ & $-$ \\
  1554+215$^{\rm a}$ & $26\,300$  & $15.2$ & $7.26$ & $105\pm6$ & $63\pm4$ & $-$ & $-$ & $-$ & MS, 6 \\
  1555--089$^{\rm a}$ & $14\,600$ & $14.8$ & $8.30$ & $220\pm13$ & $134\pm8$ & $-$ & $-$ & N & MS, 18 \\
  1601+581 & $14\,700$ & $13.8$ & $8.27$ & $520\pm30$ & $330\pm10$ & $-$ & $-$ & Y & $-$ \\
  1609+044 & $29\,100$ & $15.2$ & $7.08$ & $96\pm5$ & $59\pm3$ & $-$ & $-$ & N & $-$ \\
  1614--128 & $16\,600$ & $15.0$ & $8.01$ & $161\pm8$ & $104\pm5$ & $-$ & $-$ & N & $-$ \\
  1614+136 & $22\,000$ & $15.2$ & $7.23$ & $113\pm6$ & $68\pm4$ & $54\pm5$ & $16\pm6$ & $-$ & $-$ \\
  1619+123$^{\rm a}$ & $16\,400$ & $14.6$ & $8.03$ & $230\pm10$ & $149\pm8$ & $-$ & $-$ & N & MS, 3 \\
  1620--391$^{\rm a}$ & $24\,700$ & $11.0$ & $7.38$ & $5000\pm200$ & $3100\pm200$ & $1900\pm100$ & $1070\pm50$ & $-$ & MS, 6 \\
  1620+260 & $28\,300$ & $15.6$ & $7.15$ & $76\pm4$ & $46\pm2$ & $-$ & $-$ & N & $-$ \\
  1633+676 & $23\,700$ & $16.3$ & $7.53$ & $34\pm2$ & $22\pm1$ & $-$ & $-$ & N & $-$ \\
  1641+387 & $15\,600$ & $14.6$ & $8.31$ & $250\pm10$ & $156\pm8$ & $-$ & $-$ & Y & $-$ \\
  1643+143$^{\rm c}$ & $26\,800$ & $15.4$ & $7.26$ & $5700\pm300$ & $3800\pm200$ & $-$ & $-$ & $-$ & MS, 19 \\
  1647+375 & $22\,000$ & $14.9$ & $7.59$ & $138\pm7$ & $86\pm4$ & $-$ & $-$ & Y & $-$ \\
  1713+332 & $22\,100$ & $14.4$ & $7.32$ & $-$ & $185\pm9$ & $-$ & $57\pm6$ & Y & $-$ \\
  1713+695 & $15\,900$ & $13.2$ & $8.26$ & $820\pm40$ & $520\pm30$ & $-$ & $-$ & N & $-$ \\
  1739+804 & $26\,500$ & $15.6$ & $7.45$ & $82\pm4$ & $50\pm3$ & $-$ & $-$ & N & $-$ \\
  1755+194 & $24\,400$ & $16.0$ & $7.46$ & $55\pm3$ & $34\pm2$ & $-$ & $-$ & Y & $-$ \\
  1914--598 & $19\,800$ & $14.4$ & $7.88$ & $290\pm10$ & $200\pm10$ & $-$ & $-$ & Y & $-$ \\
  1919+145 & $15\,300$ & $13.0$ & $8.40$ & $1120\pm60$ & $690\pm30$ & $1180$ & $741$ & N & $-$ \\
  1943+163 & $19\,800$ & $14.0$ & $7.71$ & $-$ & $230\pm10$ & $-$ & $101\pm8$ & Y & $-$ \\
  1953--715 & $19\,300$ & $15.1$ & $7.96$ & $133\pm7$ & $84\pm4$ & $-$ & $-$ & Y & $-$ \\
  2009+622$^{\rm c}$ & $26\,500$ & $15.3$ & $7.15$ & $1330\pm70$ & $900\pm50$ & $-$ & $-$ & Y & MS, 20\\
  2018--233 & $15\,700$ & $15.1$ & $8.27$ & $170\pm9$ & $111\pm6$ & $-$ & $-$ & N & $-$ \\
  2021--128 & $20\,800$ & $15.2$ & $7.78$ & $113\pm6$ & $74\pm4$ & $-$ & $-$ & N & $-$ \\
  2032+188 & $18\,200$ & $15.3$ & $7.58$ & $125\pm6$ & $81\pm4$ & $-$ & $-$ & N & $-$ \\
  2039--682 & $17\,100$ & $13.3$ & $8.51$ & $-$ & $490\pm30$ & $-$ & $160\pm10$ & N & $-$ \\
  2039--202 & $19\,700$ & $12.3$ & $7.82$ & $-$ & $1040\pm50$ & $-$ & $370\pm20$ & $-$ & $-$ \\
  2043--635 & $27\,700$ & $15.6$ & $7.89$ & $72\pm4$ & $49\pm3$ & $-$ & $-$ & N & $-$ \\
  2046--220 & $23\,400$ & $15.4$ & $7.45$ & $89\pm5$ & $56\pm3$ & $-$ & $-$ & Y & $-$ \\
  2056+073 & $27\,300$ & $15.4$ & $7.75$ & $80\pm4$ & $53\pm3$ & $-$ & $-$ & N & $-$ \\
  2058+181 & $17\,400$ & $15.2$ & $8.07$ & $149\pm8$ & $97\pm5$ & $-$ & $-$ & Y & $-$ \\
  2115+010 & $25\,200$ & $15.6$ & $7.30$ & $70\pm4$ & $43\pm2$ & $-$ & $-$ & $-$ & $-$ \\
  2134+218 & $18\,000$ & $14.5$ & $8.06$ & $-$ & $159\pm8$ & $-$ & $61\pm7$ & Y & $-$ \\
  2143+353 & $26\,100$ & $15.9$ & $7.38$ & $51\pm3$ & $35\pm2$ & $-$ & $-$ & N & $-$ \\
  2149+021 & $17\,900$ & $12.8$ & $8.07$ & $-$ & $780\pm40$ & $-$ & $270\pm10$ & $-$ & $-$ \\
  2152--045 & $19\,800$ & $15.7$ & $7.44$ & $84\pm4$ & $53\pm3$ & $-$ & $-$ & Y & $-$ \\
  2200--136$^{\rm c}$ & $25\,800$ & $15.3$ & $7.19$ & $107\pm5$ & $65\pm3$ & $-$ & $-$ & $-$ & WD, 2 \\
  2204+070 & $25\,600$ & $15.8$ & $7.49$ & $70\pm5$ & $43\pm2$ & $-$ & $-$ & $-$ & $-$ \\
  2204+071 & $24\,500$ & $15.8$ & $7.34$ & $69\pm4$ & $42\pm2$ & $-$ & $-$ & $-$ & $-$ \\
  2205--139 & $26\,000$ & $15.0$ & $7.69$ & $120\pm6$ & $72\pm4$ & $-$ & $-$ & N & $-$ \\
  2210+233 & $23\,200$ & $15.8$ & $7.97$ & $60\pm3$ & $35\pm2$ & $-$ & $-$ & N & $-$ \\
  2218--271 & $15\,000$ & $14.7$ & $8.14$ & $350\pm20$ & $260\pm10$ & $-$ & $-$ & N & $-$ \\
  2220+133 & $22\,600$ & $15.6$ & $8.03$ & $75\pm4$ & $46\pm2$ & $-$ & $-$ & N & $-$ \\
  2220+217A$^{\rm a}$ & $18\,700$ & $15.5$ & $8.29$ & $80\pm4$ & $46\pm2$ & $-$ & $-$ & $-$ & WD, 2 \\
  2220+217B$^{\rm a}$ & $14\,600$ & $15.2$ & $8.13$ & $73\pm4$ & $43\pm2$ & $-$ & $-$ & $-$ & WD, 2 \\
  2225+219 & $26\,000$ & $15.8$ & $7.30$ & $60\pm3$ & $36\pm2$ & $-$ & $-$ & Y & $-$ \\
  2226+061 & $15\,300$ & $14.8$ & $7.93$ & $200\pm10$ & $126\pm6$ & $73\pm7$ & $41$ & N & $-$ \\
  2229+235 & $19\,300$ & $16.0$ & $8.03$ & $59\pm3$ & $39\pm2$ & $24\pm2$ & $-$ & Y & $-$ \\
  2231--267 & $21\,600$ & $15.0$ & $7.79$ & $162\pm8$ & $109\pm5$ & $-$ & $-$ & Y & $-$ \\
  2238--045 & $17\,500$ & $16.9$ & $8.39$ & $23\pm1$ & $13\pm1$ & $-$ & $-$ & N & $-$ \\
  2244+210 & $24\,100$ & $16.5$ & $7.41$ & $35\pm2$ & $21\pm1$ & $-$ & $-$ & Y & $-$ \\
  2248--504$^{\rm c}$ & $16\,300$ & $15.0$ & $8.19$ & $159\pm8$ & $100\pm5$ & $-$ & $-$ & N & WD, 4 \\
  2257+162$^{\rm c}$ & $24\,600$ & $16.0$ & $7.24$ & $420\pm20$ & $290\pm10$ & $200\pm10$ & $128\pm8$ & Y & MS, 9 \\
  2306+124 & $20\,400$ & $15.1$ & $7.87$ & $124\pm6$ & $77\pm4$ & $-$ & $-$ & Y & $-$ \\
  2312--356 & $15\,100$ & $15.5$ & $8.28$ & $126\pm6$ & $83\pm4$ & $-$ & $-$ & $-$ & $-$ \\
  2318--226 & $29\,900$ & $16.1$ & $7.15$ & $40\pm2$ & $24\pm1$ & $-$ & $-$ & $-$ & $-$ \\
  2322--181 & $21\,700$ & $15.3$ & $7.57$ & $100\pm5$ & $62\pm3$ & $-$ & $-$ & Y & $-$ \\
  2328+107 & $21\,000$ & $15.8$ & $7.74$ & $116\pm6$ & $77\pm4$ & $-$ & $-$ & N & $-$ \\
  2331+290 & $27\,300$ & $15.7$ & $6.96$ & $69\pm4$ & $43\pm2$ & $29$ & $14$ & $-$ & $-$ \\
  2345--481 & $29\,400$ & $15.5$ & $6.86$ & $54\pm3$ & $32\pm2$ & $-$ & $-$ & N & $-$ \\
  2345+304 & $29\,100$ & $16.4$ & $7.06$ & $43\pm2$ & $27\pm2$ & $-$ & $-$ & N & $-$ \\
  2347--192 & $27\,400$ & $16.1$ & $7.23$ & $41\pm2$ & $23\pm1$ & $-$ & $-$ & $-$ & $-$ \\
  2350--248 & $29\,700$ & $15.4$ & $7.54$ & $77\pm4$ & $50\pm3$ & $-$ & $-$ & Y & $-$ \\
  2359--324$^{\rm c}$ & $22\,500$ & $16.3$ & $7.38$ & $43\pm2$ & $28\pm2$ & $-$ & $-$ & N & WD, 4
  \label{tab:TABA1}
\end{longtable}
\end{ThreePartTable}


\bsp	
\label{lastpage}
\end{document}